%% file: 1Main.tex
\documentclass[reprint,
amsmath,amssymb,
aps,
floatfix,
]{revtex4-2}

\usepackage{graphicx}
\usepackage{dcolumn}
\usepackage{bm}
\usepackage{physics}

\begin{document}

\preprint{APS/123-QED}

\title{Spin-to-Charge conversion with electrode confinement in diamond}

\author{Liam Hanlon}%
\affiliation{%
Laser Physics Centre, Research School of Physics, Australian National University, Canberra, ACT, Australia, 2601
}%
\author{Lachlan Oberg}%
\affiliation{%
Laser Physics Centre, Research School of Physics, Australian National University, Canberra, ACT, Australia, 2601
}%
\author{Yun Heng Chen}%
\affiliation{%
Laser Physics Centre, Research School of Physics, Australian National University, Canberra, ACT, Australia, 2601
}%
\author{Marcus W. Doherty}%
 \email{mwdoherty@anu.edu.au}
\affiliation{%
Laser Physics Centre, Research School of Physics, Australian National University, Canberra, ACT, Australia, 2601
}%

\date{\today}

\begin{abstract}
The nitrogen-vacancy (NV) center in diamond has a wide range of potential applications in quantum metrology, communications and computation. The key to its use lies in how large the optical spin contrast is and the associated fidelity of spin state readout. In this paper we propose a new mechanism for improving contrast with a spin-to-charge protocol that relies on the use of an external electrode and cryogenic temperatures to discretise the diamond conduction band for spin-selective resonant photoionization. We use effective mass theory to calculate the discrete eigenenergies in this new system and use them to formulate a new spin-to-charge protocol that involves resonant photoionization out the NV ground state into the diamond conduction band. The major sources of broadening are also addressed which guide the design of the experiment. With this mechanism we theorise an optical spin contrast that and an associated spin readout fidelity of 85\%. This significant improvement can be applied to a number of cryogenic quantum technologies.      

\end{abstract}

\maketitle

The nitrogen vacancy (NV) center in diamond is a unique and well-studied defect which has been used in a variety of quantum metrology and imaging experiments \cite{Barry2016b, Kucsko2013, LeSage2013OpticalCells, Barson2021NanoscaleSpin} as well as quantum computing and networking experiments \cite{Maurer2012Room-temperatureSecond, Wu2019AConditions, Nizovtsev2005ASpins}. The defect possesses the longest electron coherence time of any solid-state defect \cite{Balasubramanian2009}, permits optical initialisation and readout of spin-states under ambient and cryogenic conditions \cite{Doherty2013TheDiamond}, and can be engineered within a myriad of nanoscale geometries \cite{Babinec2010a, Jamali2014a, Wan2018}.

The key mechanism involved in both NV quantum sensing as well as computing is the ability to initialise and readout the NV electron spin state. The electronic structure that allows for spin selective readout is only available when the NV center is in the negatively charged state (NV$^-$), other charge states do not exhibit spin selective properties. This presents a significant problem in NV physics as the center can ionize into the neutral NV$^0$ state under optical illumination where the electron is excited into the diamond conduction band. During this process the spin information is lost and the NV center isn't usable until an electron repopulates the defect, converting it back into NV$^-$. The recombination process is most often achieved by optically exciting electrons into the NV center from the diamond valence band\cite{Aslam2013Photo-inducedDetection}. This ionization process limits NV charge state control and is an important issue in NV physics. 

One key area of study for NV charge state control involves purposeful photoionization via the spin-to-charge conversion (SCC) technique. With SCC, the NV spin information is read out by optically ionizing the defect when it is in a particular spin state. Thus, the spin of the electron is mapped to the charge state of the NV and spin information is obtained by measuring the charge state. The charge state can be measured optically with a laser maximally resonant to NV$^-$\cite{Jaskula2019ImprovedConversion} or by measuring the photoelectric current induced by ionization \cite{Siyushev2019PhotoelectricalDiamond, Gulka2021Room-temperatureSpins}. This technique has been shown to have a larger optical spin contrast compared to conventional intrinsic photo-luminescence cycling techniques \cite{Siyushev2019PhotoelectricalDiamond, Jaskula2019ImprovedConversion, Zhang2021High-fidelityConversion}, which in turn increases the readout fidelity of measuring a spin state. Whilst SCC methods improve contrast, the improvement is not very significant. For example, Jaskula et al reported an SCC readout contrast of 36\% compared to the conventional methods which have a readout contrast of 25\% \cite{Jaskula2019ImprovedConversion}. The primary reason for the limitation is that SCC protocols still require shelving the NV electron into the singlet state manifold via an inter-system crossing (ISC). The branching rate of the ISC is not 100\% from the $m_s=\pm1$ state nor is it 0\% from the $m_s=0$ state \cite{Goldman2015Phonon-inducedCenters}, this lowers the spin-selectivity and increases the probability of a false readout. Current methods also do not alter the probability of photoionization itself, the rate of photoionization is set by the intrinsic nature of the NV center in diamond. Being able to read out an electron spin state consistently without introducing noise from erroneous photoionization is essential in producing high fidelity spin measurements for quantum protocols.

We introduce a new mechanism of charge state control with the application of an external electrode to the surface of the diamond over a near surface NV center. The electrode creates a potential well within the diamond which has the effect of spectrally confining the density of low lying conduction band states in the diamond. This discretized conduction band has a two-fold effect. Firstly, it increases the photoionization probability at frequencies resonant to a discrete transition whilst reducing the probability of photoionization at other frequencies. Secondly, the electrode creates energy level separation in the conduction band which is much larger than the separation of levels in the NV ground state triplet. These two factors allow for an SCC protocol where the NV electron is resonantly ionized out of the ground state triplet into a discrete conduction band state with higher probability compared to conventional photoionization. The wide separation of the conduction band states means that the individual triplet transitions can be addressed. These factors create a highly selective spin to charge protocol with very high optical spin contrast. The technique promises to vastly improve the fidelity of spin readout which has applications for NV based quantum sensing, communications and computation. The design also creates a discrete three level system for stimulated Raman adiabatic passage (STIRAP) experiments \cite{Oberg2019SpinDiamond}. The key to this process is ensuring that the spectral lines are narrow enough to be resolved. To achieve this the major sources of broadening need to be addressed. 

In this paper we provide a proof-of-principle demonstration of the electrode for NV readout using extensive theoretical modelling. We introduce the electrode and briefly describe how its design improves performance. Simulations of the electric potential well being generated by the electrode are performed and effective mass theory is then applied to calculate the discrete eigenenergies and wavefunctions being produced in the well. We then use these energies to calculate the density of states and subsequent photoionization probability. Next, we consider the major sources of broadening which describe the constraints on the design of the system and help outline problems to be overcome to fully realise this technology. Finally, we describe the spin to charge conversion protocol, outline the requirements of the system for the protocol to be successful, and calculate the contrast and readout fidelity this protocol can produce.  

\begin{figure}[!ht]
    \centering
    \includegraphics[width = 0.50\textwidth]{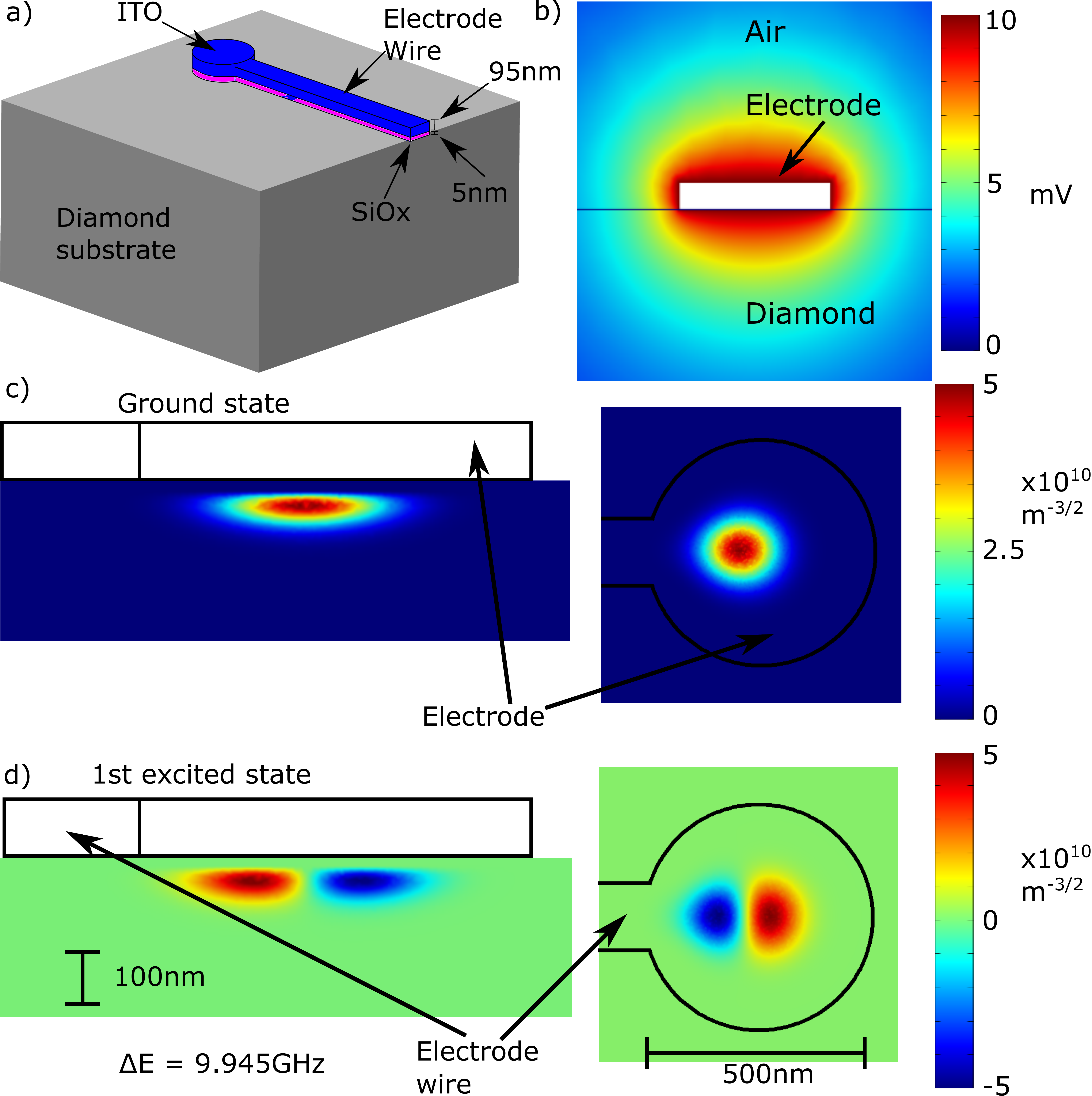}
    \caption{a) Image of the electrode setup. The grey diamond has an electrode fabricated onto it made of 5~nm SiOx and 95~nm ITO. The cylindrical section of the electrode sits over an NV center and a wire connects the electrode to a voltage source. b) Simulation of the potential well generated by the electrode. The white section is an electrode which sits on a diamond surface and carries a 10~mV potential. c) Plot of the simulated wavefunction for the ground state wavefunction viewed from the XZ (left) and XY plane (right image). d) Plot of the same simulation from c) but for the first excited state wavefunction. The $\Delta E$ term is the difference between the first and second eignenergies.}
    \label{fig:electrode}
\end{figure}

In our design, the cylindrical electrode is placed over an NV with a thin wire connecting to a voltage source. The electrode and wire have a thin, insulating silicon oxide (SiOx) layer to prevent charges moving from the diamond into the electrode. On top of the SiOx is a transparent indium-tin oxide (ITO) conductive layer which carries the electric potential and allows for optical illumination through the electrode. The dimensions of the electrode were chosen to maximise the energy splitting in the diamond conduction band (see figure \ref{fig:electrode}). Smaller confinement volumes (from smaller electrodes) have larger energy splitting, making them easier to individually address. The limiting factor on size comes from the resolution of the nanofabrication process itself as the wire width must be smaller than the electrode to prevent wavefunctions occupying the space under the wire (see supplementary for more details). 

The potential well generated by the electrode can be solved for using Maxwell's equations. The solution can then be used in a Schrodinger equation which solves for the required eigenenergies and associated wavefunctions: 

\begin{equation}\label{eqn:schrodinger}
    \Big(-\frac{\hbar^2}{2} \Vec{\nabla} \cdot \overleftrightarrow{\frac{1}{m}} \cdot \Vec{\nabla} + V(\vec{r})\Big)\ket{F_n(\vec{r})} = E_n\ket{F_n(\vec{r})},
\end{equation}

where $E_n$ is the eigenenergy for a given energy level $n$, $V(\vec{r})$ is the electric potential from the electrode, $\overleftrightarrow{m}$ is the effective mass tensor of the electron and $F_n$ is an envelope function. The envelope function is related to the electron wavefunction by the following:

\begin{equation}\label{eqn:envelope}
   \psi_n = F_n(\vec{r}) u_{\vec{k}}(\vec{r}) e^{i \vec{k} \cdot \vec{r}}, 
\end{equation}

where $u_{\vec{k}}(\vec{r})$ is the Bloch function for the state of the conduction band minimum in the bulk diamond unit cell and $\vec{k}$ is its associated wavevector. A full derivation of equation \ref{eqn:schrodinger} can be found in the supplementary section. 

These equations are applicable due to three key assumptions that we apply in effective mass theory. The first is that for photoionization we only need to consider states that are energetically close to the conduction band minimum (CBM). This is due to the fact that we are only ionizing from the NV to the lowest lying states in the conduction band. This assumption allows us to approximate the electron energy as a free electron with an effective mass in a confining potential. The second assumption is that the envelope function varies on a distance much greater than the Bloch function, allowing it to be considered constant when calculating values over the Bloch function space. Finally, we assume that the electrode potential isn't strong enough or varying enough on the scale of the unit cell such that the Bloch function depends on the potential. 

Stronger electrode potentials also reduce the depth of the confining wavefunction, requiring the placement of NVs closer to the diamond surface in order to be effected by the electron confinement. This is undesirable as NVs close to the diamond surface exhibit charge instability\cite{DeOliveira2017}. The potential chosen was designed to create a wavefunction whose center is at the position of the NV which is approximately 50~nm from the diamond surface. This choice was made to maximise the eigenenergy splitting whilst maintaining NV stability. 

Three separate solutions from equation \ref{eqn:schrodinger} were obtained to account for the three effective masses in the separate Cartesian directions along the diamond. In each solution, 6 eigenenergies/wavefunctions were obtained and added together to give a total of 18 conduction band energy levels. The eigenenergies calculated were then be used to calculate the transition rate from the NV to a particular conduction band state by using Fermi's golden rule:

\begin{equation}\label{eqn:blochEnvelope}
    \Lambda_{n}(E) = \frac{2\pi}{\hbar} \sum_n |\mu_{b} C(E) \sqrt{A_e}F_{n}(r)|_{r=NV} \mathcal{E}(E)|^2 \mathcal{L}(E-E_n), 
\end{equation}

where $\mathcal{L}(E-E_n)$ is a Lorentzian function whose peaks are at the energy levels of the conduction band states, $E_n$, $\mu_{b}$ is the transition dipole moment in bulk diamond, $\mathcal{E}(E)$ is the electric field of the interrogating laser and $C(E)$ is the dimensionless Franck-Condon factor which describes the vibrational overlap of states under a Born-Oppenheimer approximation. The electron confinement envelope function is represented by $F_{n}(r)$ which is defined at the position of the NV and is normalised to the volume of the diamond with the constant $A_e$.

\begin{figure}[!ht]
    \centering
    \includegraphics[width = 0.45\textwidth]{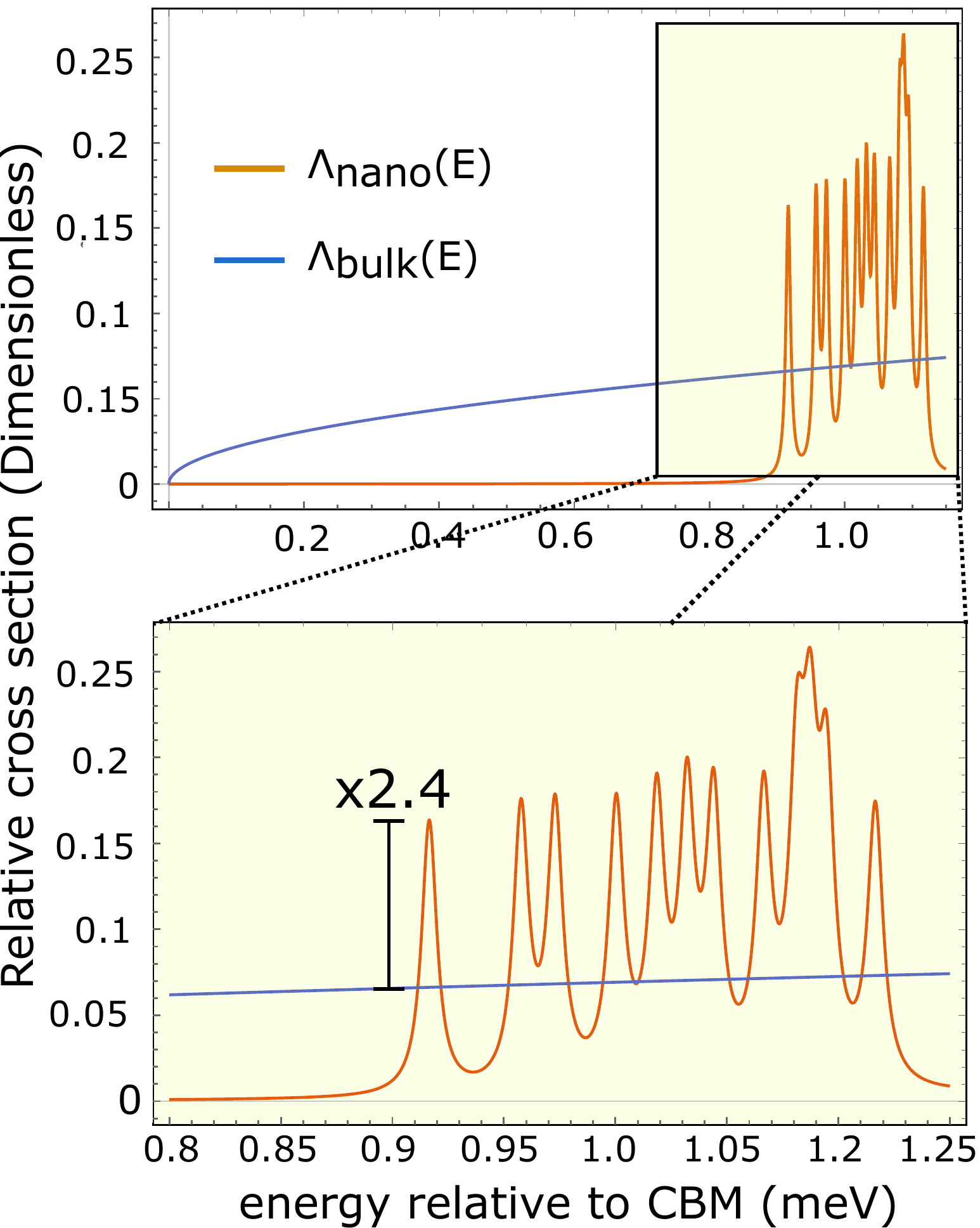}
    \caption{Plot of the transition cross section in both bulk diamond (blue) and in the confined region created by the electrode (orange). The cross section is proportional to $\sqrt{E}$ in bulk diamond whereas the confined electron has Lorentzian peaks at each eigenenergy calculated using equation \ref{eqn:schrodinger}. The linewidth: $\phi=1$~GHz, is chosen based on the error modelling and shows how thin lines are easily distinguishable. Taking the ratio of the first peak with the equivalent bulk value gives a transition rate that is 2.4 times more likely for a confined electron.}
    \label{fig:transition}
\end{figure} 

Figure \ref{fig:transition} shows the transition cross section for photoionization in a confined electrode (orange) and bulk diamond (blue) for the first 12 energy levels. The plots are normalised to the dipole moment, electric field and Franck-Condon factor which is common to both transition cross sections. The equations have been put in dimensionless units such that the ratio of photoionization between the two curves can be assessed (see supplementary for details).  Whilst the bulk diamond transition rate is a smooth function proportional to $\sqrt{E}$, the transition rate in the confined system shows clear Lorentzian peaks at each eigenenergy calculated using equation \ref{eqn:schrodinger}. When the electron is confined, the probability of resonant ionization to the first conduction band level is 2.4 times higher compared to bulk diamond, increasing charge state control.  

Distinguishing between adjacent transitions is a key aspect of the SCC protocol. Broadening of the transition lines will increase the probability of an unwanted transition and will prevent accurate readout of the NV spin state. From figure \ref{fig:transition}, the transition linewidths are approximated to be about 1~GHz and the transitions are qualitatively distinguishable, however this needs to be considered in more detail. The three main sources of linewidth broadening are identified as: fluctuations in the confining potential from the electrode, fluctuations in the confining potential due to surface charge traps and fluctuations in the conduction band energy levels from electron-phonon broadening. 

Inconsistencies in the electrode potential will alter the potential at the NV and the resultant diamond conduction band states. This source of broadening was assessed using the following equation: 

\begin{equation}\label{eqn:electrode}
    \Gamma_{electrode} = \sigma_V\Big(\frac{\partial E_n}{\partial V}-\frac{1}{h} \frac{\partial E_{NV}}{\partial V} \Big),
\end{equation}

where $\sigma_V$ is the RMS uncertainty of the electrode which was assumed to be $\pm 0.001mV$, $\frac{\partial E_n}{\partial V}$ is the change in conduction band energy as a function of potential (GHz/mV) and $\frac{\partial E_{NV}}{\partial V}$ is the change in NV energy level as a function of potential. The change in conduction band energy can be calculated by solving for the eigenenergies with an offset to the confining potential by $\pm 0.001mV$ and calculating the slope (see supplementary for more details and figures). The change in NV energy is given by $-e V$ where $e$ is the electron charge. Putting all these factors together gives a linewidth of 0.257~GHz.

Electron-phonon (e-p) scattering is the process by which the energy of an electron is altered slightly by a phonon interaction. In bulk diamond this can be an issue even at low temperatures (4~K). We can model this broadening with Fermi's golden rule: 

\begin{equation}\label{eqn:e-p}
    \begin{split}
    \Gamma_{ep} = \alpha\sum_n \int |\bra{n} \Psi_k (\vec{r}) \ket{1}|^2 \omega_k^3 n_B&(\omega_k) \delta(\omega_n - \omega_k)d^{3}k\\
    \alpha = \frac{\Theta^2}{2(2\pi)^2\hbar\rho c_{l}^4},
    \end{split}
\end{equation}

where $\Theta$ is the acoustic deformation potential, $\rho$ is the density of diamond and $c_l$ is the longitudinal speed of sound in diamond. The equation is a sum of all the interactions of the phonon modes $\Psi_k(\vec{r})$ with the discretised energy levels of the conduction band minimum from 1 to n, integrated over the phonon k-space. The phonons have a wavelength $\omega_k$ and a temperature dependent distribution given by a Bose-Einstein distribution $n_B$ \cite{Oberg2019SpinDiamond}. The full derivation of this equation is given in the supplementary material. For simplicity of calculation, the confining potential is modelled as a square well and then solved for a range of volumes (plots available in the supplementary). For a square well that has roughly the same dimensions as the electrode confining potential volume, the e-p broadening does not exceed 0.5~GHz.

Imperfections in the diamond structure as well as its surface termination allow for surface charge traps that can hold electrons \cite{Sasama2020Charge-carrierTransistors}. Surface charge traps can be occupied briefly by electrons both from the NV as well as from defects in the bulk diamond. Charges from these surface traps can then ionize back into the bulk diamond or can hop from trap to adjacent trap. The result is a continuous fluctuation of a local electric field from the constant change in the position of the charges relative to the NV. The noise from the electric field fluctuation can affect both the NV and conduction band energy similar to noise from the electrode. To calculate the broadening we assume that the surface charge traps are uniformly distributed on the diamond and that the electrons occupying the traps can only move from trap to adjacent trap. In this picture, the charge motion can be modelled using Redfield theory for a two level fluctuator \cite{Silchter1996PrinciplesResonance} where the linewidth is given by:

\begin{equation}\label{eqn:surfacecharge}
    \Gamma_{sc} = \frac{2\pi}{\hbar^2} \overline{|\delta E|^2} S(0),
\end{equation}

where $S(\omega)$ is the Lorentzian noise spectrum and $\overline{|\delta E|^2}$ is the variance of the change in energy which is given by:

\begin{equation}\label{eqn:deltaE}
    \overline{|\delta E|^2} = \eta e^2 \int_{\Omega} |\bra{F_0}V(\vec{r},\vec{q})\ket{F_0}-V(\vec{r}_{NV}, \vec{q})|^2 d^2q,
\end{equation}

where $\eta$ is the surface density of acceptors and $V(\vec{r},\vec{q})$ is the potential generated by an electron at position $\vec{q}$ on the surface of the diamond away from the NV at position $\vec{r}$. A full derivation of this broadening can be found in the supplementary material. By treating the ground state conduction band envelope function, $F_0$, as a Gaussian then the effect of a single surface trap on the Gaussian distribution can be modelled as a multipole expansion. The solution of a single expansion can then be integrated over the number of surface traps to obtain the overall electric field noise and its associated broadening. When solving for the overall linewidth using this model the broadening is actually significant. Even when considering extremely fast charge motion on the surface (THz in order) the broadening can be as high as $10^{15}$~Hz for a typical trap density ($10^{18}$~m$^{-2}$)\cite{Oberg2020SolutionElectrometers, Stacey2019EvidenceSources}, as much as 6 orders of magnitude broader than the 1~GHz broadening shown in figure \ref{fig:transition}. 

The ideal way to mitigate the surface trap issue is to reduce the density of charges on the surface of the diamond. Whilst physically reducing the surface trap density is difficult (although not impossible), an alternate solution is to fill the traps with a nitrogen donor layer in the diamond which is considered in more detail by Oberg et. al.\cite{Oberg2020SolutionElectrometers}. The donor layer nitrogen's will pass their electron to the surface traps, resulting in a complete filling of the traps. The filling of the traps will mitigate any opportunity for electrons to hop within the traps, thereby reducing the AC electric field they might produce. The amount of trap filling that occurs depends on the concentration of donors relative to the trap concentration as well as the distance the donors are from the traps. Calculating the occupation of the traps can be achieved by defining the trap energy and density in a Fermi-Dirac distribution (see supplementary for full derivation). Full surface trap occupation can be achieved by adding a donor nitrogen layer 100~nm from the surface of the diamond and reducing the trap density to $\approx10^{15}$~m$^{-2}$ with an equal concentration of donors. Fully occupied traps allow no change in the surface electric field from charge motion as all the traps are filled, thereby reducing the broadening from surface traps to effectively zero. 

\begin{figure}[!ht]
    \centering
    \includegraphics[width = 0.4\textwidth]{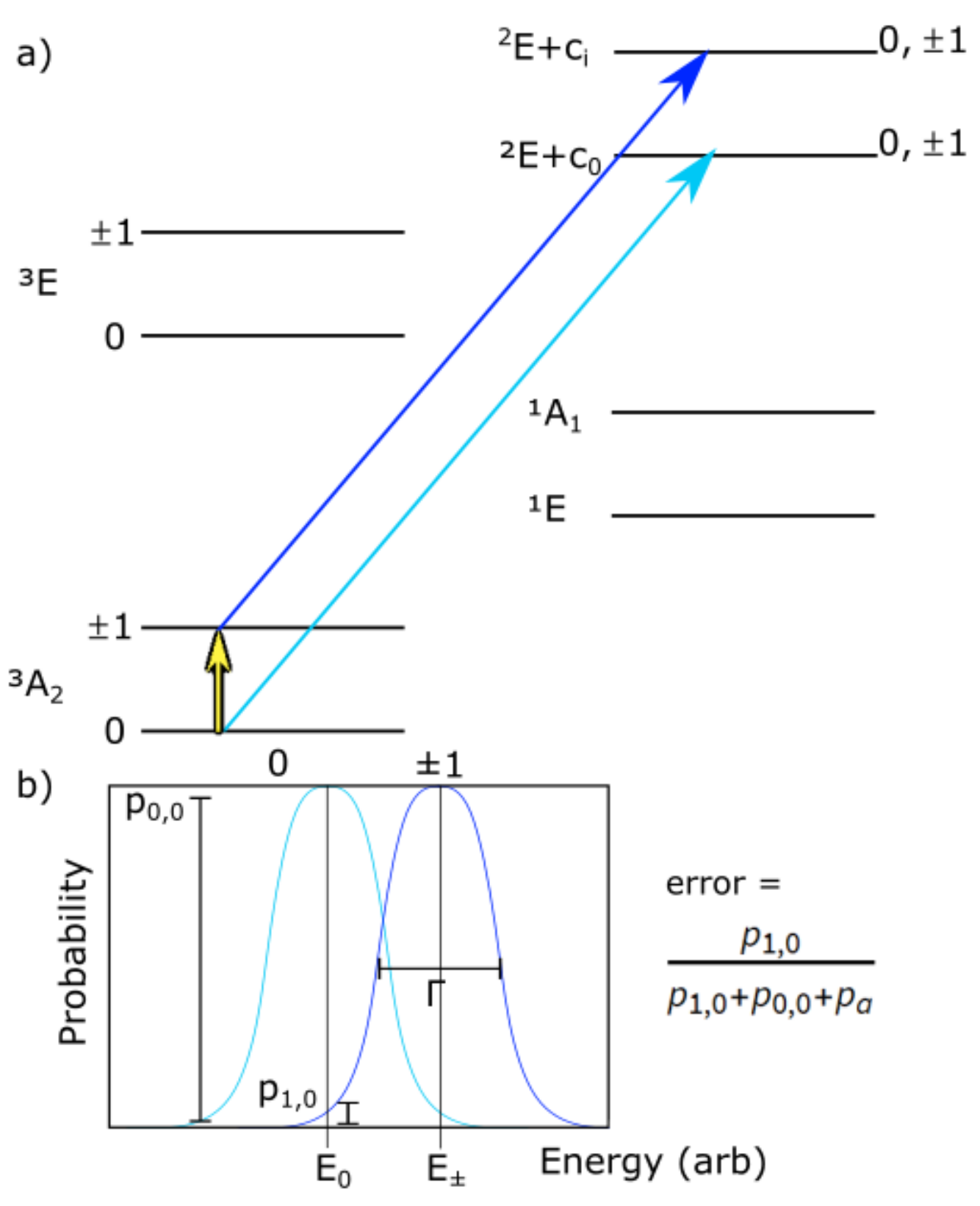}
    \caption{a) Diagram of the spin to charge readout protocol. With the electrode in place, the low lying conduction band states are discretized. This allows for resonant ionization into the upper energy states which represent the ionized NV$^{0}$ ${}^2$E state with an electron in one of the first two conduction band levels $c_{0}$ and $c_{i}$ respectively. Thus a laser (light blue) can be applied to ionize out of the ground state $m_{s}=0$ and a microwave/laser (yellow/dark blue respectively) combination can be used to ionize out of the $m_{s}=\pm 1$. b) As long as the linewidth $\Gamma$ is small and the energy difference between the two transitions is large then the two transitions can be distinguished. In the error equation, $p_{1,0}$ is the probability of ionizing from the wrong spin state when intending ionization from the other, $p_{0,0}$ is the probability of ionizing correctly from the correct spin state and $p_a$ is the probability of absorbing instead of ionizing. Thus, the error is the ratio of erroneous photoionization compared to the total amount probability of other processes (see supplementary for more details).}
    \label{fig:SpinToCharge}
\end{figure} 

With the electrode and diamond design characteristics outlined, the spin-to-charge protocol is relatively simple. When the NV electron is in the ground triplet, an ionizing laser excites the electron from the ground state to the upper energy states which represent the ionized NV$^0$ ${}^2$E state with an electron in one of the first two conduction band levels $c_0$ and $c_i$ respectively (figure \ref{fig:SpinToCharge}). The NV$^0$ charge state can then be read out either optically or through measuring the photoelectric current \cite{Jaskula2019ImprovedConversion, Siyushev2019PhotoelectricalDiamond}. The ionization will present as a drop in fluorescence from the NV$^-$ or by an increase in current as the NV is ionized into the neutral charge state. This process can occur when ionizing from $m_s=0$ to the conduction band or $m_s=\pm 1$ to the conduction band. Fine structure of the first two conduction band states is negligible due to the fact that there is no spin-orbit effects in the conduction band minimum \cite{Oberg2019SpinDiamond}. 

To calculate the overall optical readout contrast the following equation is applied: 

\begin{equation}\label{eqn:contrast}
    C = \frac{1 - \mathcal{L}(\Delta E)}{1 + \mathcal{L}(\Delta E) + g/f},
\end{equation}

in the above equation, $g$ is a constant factor that describes the ratio of photoionization to absorption at the expected photoionization energy (2.6~eV \cite{Aslam2013Photo-inducedDetection}) which is taken from Razinkovas et al \cite{Razinkovas2021VibrationalDiamond}. The $f$ term is a constant that describes the increase in photoionization cross section due to the discretization of the conduction band minimum taken from figure \ref{fig:transition}. Finally,   $\mathcal{L}(\Delta E)$ is the Lorentzian function that describes the photoionization spin selectivity. The splitting of the NV ground state triplet ($m_s=0$ and $m_s=\pm 1$ respectively) is known to be: $\Delta D \approx 2.87$~GHz \cite{Manson2018NV--N+Diamond, Doherty2013TheDiamond} and the splitting of the first two conduction band levels under the electrode is calculated to be: $\Delta C \approx 9.945$~GHz using equation \ref{eqn:schrodinger}. The Lorentzian is then a function of the difference in energy separation between the conduction band energy levels and the ground state NV triplet energy levels ($\Delta E = \Delta C - \Delta D = 7.075$~GHz) as well as the total linewidth broadening which is found by adding the broadening sources considered in this paper ($0.757$~GHz). Equation \ref{eqn:contrast} is derived in more detail in the supplementary section and gives an optical spin contrast of 85\%. The associated spin readout fidelity of the readout can be expressed with the following equation:

\begin{equation}\label{eqn:fidelity}
    F = 1-\Big(\frac{\mathcal{L}(\Delta E) + g/f}{1 + \mathcal{L}(\Delta E) + g/f}\Big), 
\end{equation}

where a full derivation can be found in the supplementary section. From this equation the readout fidelity can be calculated to also be 85\%.  

Readout fidelity is one of the main limiting factors in NV performance for a variety of quantum technologies. Typically in the NV, spin readout occurs when the electron is pumped into the singlet levels from the triplet manifold via an ISC. The branching ratio in the ISC reduces the probability of electrons transitioning to the singlet from the $m_s=\pm 1$, reducing overall contrast \cite{Goldman2015Phonon-inducedCenters}. This SCC protocol avoids this issue by performing a single transition photoionization that avoids the ISC altogether. It is only achievable with the electrode as the two spectral lines need to be distinguishable and this is only achieved with discrete, wide-gap conduction band energy levels. Achieving this high readout fidelity SCC using an external electrode is possible as long as the electrode remains stable, the diamond is at cryogenic temperatures and is engineered with a nitrogen donor layer with a surface trap density of $10^{15}$~m$^{-2}$. Future work would involve engineering the electrode, measuring confinement and then fully realising the design parameters for the SCC readout protocol. All these requirements are achievable, offering great potential for the future of NV technology. 

\begin{acknowledgments}
The authors acknowledge the support from the Australian Research Council (DP 170103098). We would also like to thank Audrius Alkauskas and Lukas Razinkovas for sharing absorption and photoionization cross-section data for our calculations. Finally we wish to thank Sophie Stearn for her assistance with proof-reading. 

\end{acknowledgments}

\include{2Appendix}

\bibliography{1Main}

\end{document}

%% file: 2Appendix.tex
\appendix

\section{Electrode fabrication}

The electrode is fabricated using a sputtering process. The diamond is cleaned using a three-acid boil (sulfuric, nitric and perchloric acid), after which approximately 100nm of polymethyl methacrylate (PMMA) is spin coated onto the diamond to be used as a resist layer. The electrode shape is then etched into the PMMA along with the electrode wire using electron beam lithography (EBL). The resist is then exposed to methyl iso-butyl ketone (MIBK) for approximately 1 minute to etch away the PMMA exposed by the EBL beam and is further exposed using a low dose oxygen plasma etch in a barrel etcher. This creates the hole in the resist layer which can be layered with 5~nm of silicon oxide (SiOx) and 95nm of indium-tin oxide (ITO) using sputter deposition. The remaining PMMA layer is removed using a lift-off technique with acetone where only the SiOx and ITO layers remain. The wire connects to the end of the diamond which can be layered coarsely with a conductive material (for example: silver paste) that can be more easily connected to a power supply for generating a potential using a wire bonding process.

The electrode was designed to carry a potential which can create the confining well within the diamond substrate around the NV. Numerical calculations revealed that smaller confining regions produced larger conduction band splitting which is necessary for the high contrast spin-to-charge readout mechanism mentioned in the main paper. The length and width of the electrode controls the length and width of the potential well whereas the depth of the well is largely controlled by the magnitude of the potential at the electrode rather than its physical height. A larger positive potential creates a smaller confining well closer to the surface of the diamond, creating the surface instability mentioned in the main paper. The SiOx was applied to create an insulating layer between the electrode and the diamond surface and the ITO was chosen in for its capacity to carry a potential as well as its optical transparency so laser illumination can occur over the diamond surface.   

\section{Effective mass theory}

In order to calculate the conduction band wavefunctions and eigenenergies which are confined by the electrode potential we apply effective mass theory. We begin by solving the Schrodinger equation in the absence of an external potential: 

\begin{equation}\label{eqn:schrodinger2}
    \Big(\frac{-\hbar^2}{2} \Vec{\nabla} \cdot \overleftrightarrow{\frac{1}{m}} \cdot \Vec{\nabla} + V_c(\vec{k})\Big) \ket{F_n(\vec{r})} = E_n^c\ket{F_n(\vec{r})},
\end{equation}

where $E_n^c$ is the eigenenergy of the crystal system for a given energy level $n$, $V_c(\vec{k})$ is the crystal potential, $m$ is the effective mass tensor of the electron and $F_n$ is an envelope function which is related to the electron wavefunction by the following:

\begin{equation}\label{eqn:envelope2}
   \psi_n = F_n(\vec{r}) u_{\vec{k}}(\vec{r}) e^{i \vec{k} \cdot \vec{r}},
\end{equation}

where $u_{\vec{k}}(\vec{r})$ is the Bloch function for the state of the conduction band minimum in the bulk diamond unit cell and $\vec{k}$ is its associated wavevector. The exponent describes the phase difference when going between unit cells and we are considering $n$ conduction band minima as we expect new minima in different vector directions. Expanding out equation \ref{eqn:schrodinger2} to include the Bloch function gives: 

\begin{equation}\label{eqn:schrodinger3}
    \Big(T(\vec{r}) + V_c(\vec{k})\Big)F_n(\vec{r}) \mu_{\vec{k}}(\vec{r}) e^{i \vec{k}_{n} \cdot \vec{r}} = E_n^c F_n(\vec{r}) \mu_{\vec{k}}(\vec{r}) e^{i \vec{k}_{n} \cdot \vec{r}},
\end{equation}

where $T(\vec{r}) = \frac{-\hbar^2}{2} \Vec{\nabla} \cdot \overleftrightarrow{\frac{1}{m}} \cdot \Vec{\nabla}$, is the kinetic energy for a free electron with an effective mass $m$. For simplicity the Bloch function can be simplified to: $\mu_{\vec{k}}(\vec{r}) e^{i \vec{k}_{n}\cdot \vec{r}} = \ket{\phi(\vec{r})}$. Expanding equation \ref{eqn:schrodinger3} as a product rule whilst multiplying both sides of the equation by the complex conjugate $\bra{\phi(\vec{r})}$ gives the following: 

\begin{equation}\label{eqn:schrodinger4}
    \begin{split}
    \braket{\phi(\vec{r})}T(\vec{r})F_n(\vec{r}) + F_n(\vec{r})\bra{\phi}T(\vec{r})\ket{\phi(\vec{r})} + \\F_n(\vec{r})\bra{\phi}V_c(\vec{k})\ket{\phi(\vec{r})} = E_n^c F_n(\vec{r})\braket{\phi},
    \end{split}
\end{equation}

note that the envelope function is not acting on the eigenstates. The envelope function varies on distances much larger than Bloch function between diamond unit cells. Thus when considering small length scales of the diamond unit cell, one approximation being made is that $F_n$ is effectively constant and can be moved out of the inner product. Taking the inner product: 
 
\begin{equation}\label{eqn:schrodinger5}
    T(\vec{r})F_n(\vec{r}) + E_b F(\vec{r}) = E_n^c F_n(\vec{r}),
\end{equation} 

where $E_b$ is the energy of the Bloch function for the conduction band minimum. Another other key approximation being made is that an electrode confining potential isn't strong enough or varying enough on the scale of the unit cell such that the Bloch function depends on the potential. This means that the electrode potential can be added in a new Schrodinger equation as the Bloch function energy is independent of the electrode potential. Adding in the electrode potential gives equation \ref{eqn:schrodinger} from the main paper:

\begin{equation}\label{eqn:schrodinger6}
    \big(T(\vec{r})+V(\vec{r})\big)F_n(\vec{r}) = E_n F_n(\vec{r}),
\end{equation}

where $E_n = E_n^c - E_b$ so that $E_n$ becomes the energy of the total wavefunction relative to the Bloch function. 

Three separate solutions from equation \ref{eqn:schrodinger} were obtained to account for the three effective masses in the separate Cartesian directions along the diamond. In each solution 6 eigenenergies/wavefunctions were obtained and added together to give a total of 18 conduction band energy levels. The eigenenergies can then be used to calculate the transition rate from the NV to a particular conduction band state by using Fermi's golden rule: 

\begin{equation}\label{eqn:blochEnvelope2}
    \Lambda_{n}(E) = \frac{2\pi}{\hbar} \sum_n |\mu_{b} C(E) \sqrt{A_e}F_{n}(r)|_{r=NV} \mathcal{E}(E)|^2 \mathcal{L}(E-E_n),
\end{equation}

where $\mathcal{L}(E-E_n)$ is a Lorentzian function whose peaks are at the energy levels of the conduction band states $E_n$, $\mu_{b}$ is the transition dipole moment in bulk diamond, $\mathcal{E}(E)$ is the electric field of the interrogating laser and $C(E)$ is the dimensionless Franck-Condon factor which describes the vibrational overlap of states under a Born-Oppenheimer approximation. The electron confinement envelope function is represented by $F_{n}(r)|_{r=NV}$ which is defined at the position of the NV and is normalised to the volume of the diamond with the constant $A_e$ using the following equation:

\begin{equation}\label{eqn:NormNano}
    \frac{A_e}{V_c}\int_V F_n^* F_n dV = 1,
\end{equation}

where $V_c$ is the volume of the diamond unit cell. Equation \ref{eqn:NormNano} is solved for $A_e$ by using the results of equation \ref{eqn:schrodinger6} and numerically integrating the envelope function over the diamond volume directly.

For comparison purposes, it is important to derive the transition rates from the NV to both the diamond conduction band in bulk as well as to the conduction band in the confining electrode potential. To do this, the first step is to calculate the density of states for both cases. With the electrode, the density of states is simply the Lorentzian function with peaks at the given energies calculated from equation \ref{eqn:schrodinger2}:

\begin{equation}\label{eqn:DOSelectrode}
    \rho_{e}(E) = \sum_n \mathcal{L}(E-E_n),
\end{equation}

where the Lorentzian function can be explicitly written as:

\begin{equation}\label{eqn:Lorentz}
    \mathcal{L}(E-E_n) = \frac{\Gamma/\pi}{(E-E_n)^2+\Gamma^2},
\end{equation}

and $\Gamma$ is the total associated linewidth. In bulk diamond it is easier to express the density of states when the energy is in terms of a wavevector, $\vec{k}$:

\begin{equation}\label{eqn:DOSbulk}
    \rho_{b}(E) = \sum_{v} \sum_{\vec{k}} \delta(E-E_{\vec{k}}),
\end{equation}

where $v$ denotes a sum over the valleys in the Bloch function. The sum of states can then be re-expressed as an integral of states over a sphere:

\begin{equation}\label{eqn:DOSbulk2}
    \rho_{b}(E) = \frac{4 \pi V}{(2\pi)^3}\int_{0}^{\infty} \vec{k}^2 \delta(E-E_{\vec{k}}) d \vec{k},
\end{equation}

by taking an effective mass argument, the energy can be expressed as the energy for a free particle with an effective mass, $m$, such that $E_{\vec{k}} = \frac{\hbar^2}{2m} \vec{k}^2$. Rearranging the equation in terms of $\vec{k}$ and substituting it into the integral gives: 

\begin{equation}\label{eqn:DOSbulk3}
    \rho_{b}(E) = \frac{V}{(2\pi)^2}\int_{0}^{\infty} \frac{2m E_{\vec{k}}}{\hbar^2} \sqrt{\frac{2m E_{\vec{k}}}{\hbar^2}} \frac{1}{2 E_{\vec{k}}} \delta(E-E_{\vec{k}}) d E_{\vec{k}},
\end{equation}

which when solved gives the following: 

\begin{equation}\label{eqn:DOSbulkfinal}
    \rho_{b}(E) = \frac{V}{2(2\pi)^2}\Big(\frac{2m}{\hbar^2}\Big)^{3/2} \sqrt{E}, 
\end{equation}

note the square root dependence on the energy, which carries into the transition rate calculation. The transition rate in the bulk diamond is then calculated using Fermi's golden rule:

\begin{equation}\label{eqn:TransitionBulk1}
    \Lambda_b = \frac{2\pi}{\hbar} \Big|\mu_{b} C(E) \sqrt{A_b}F_{b}(r) |_{r=NV}\mathcal{E}(E)\Big|^2 \sum_{\vec{k}} \delta(E-E_{\vec{k}}),
\end{equation}

where $\mu_{b}$ is the transition dipole moment in bulk diamond which is constant for all wavevectors close to the conduction band minimum and the $\sum_{\vec{k}} \delta(E-E_{\vec{k}})$ term is the density of states in bulk diamond. The normalization constant is found by integrating the envelope function over the bulk diamond volume: 

\begin{equation}\label{eqn:NormBulk}
    \frac{A_b}{V_c}\int_V F_b^* F_b dV = 1,
\end{equation}

where $V_c$ is the volume of the diamond unit cell. In bulk diamond, the envelope function encompasses all block wavefunctions so the probability of an electron existing in the envelope function will be unity: $ F_b^* F_b = 1$, therefore: $A_b = V_{c}/V$. Substituting in the normalisation and the density of states gives the following: 

\begin{equation}\label{eqn:TransitionBulkfinal}
    \Lambda_b = \frac{1}{4\pi \hbar} \Big(\frac{2m}{\hbar^2}\Big)^{3/2} \sqrt{E} \Big|\mu_b C(E)\sqrt{V_c}\mathcal{E}(E)\Big|^2,
\end{equation}

equation \ref{eqn:TransitionBulkfinal} is plotted as the blue curve in figure \ref{fig:transition} of the main paper. The transition rate in the electrode confined diamond is largely the same but with a different density of states and a different normalization: 

\begin{equation}\label{eqn:Transitionelectrode1}
    \Lambda_e = \frac{2\pi}{\hbar} \sum_n \mathcal{L}(E-E_n) \Big|\mu_n \mathcal{E}(E)\Big|^2, 
\end{equation}

where the transition dipole moment in the electrode can be expressed in terms of the bulk dipole moment mediated by the envelope wavefunction for the confined electron: $\mu_n = \mu_b C(E) \sqrt{A_e} F_e(r)|_{r-NV}$:

\begin{equation}\label{eqn:Transitionelectrodefinal}
    \Lambda_e = \frac{2\pi}{\hbar} \sum_n \mathcal{L}(E-E_n) \Big|\mu_b C(E) \sqrt{A_e} F_e(r)|_{r=NV} \mathcal{E}(E)\Big|^2,
\end{equation}

the normalization constant $A_e$ found using equation \ref{eqn:NormNano}. The energy levels ($E_n$) are calculated using equation \ref{eqn:schrodinger} from the main paper which can then be input into equation \ref{eqn:Transitionelectrodefinal} to create the orange peaked curve in figure \ref{fig:transition} of the main paper. 

The two transition rates calculated cannot be directly plotted as the dipole moment, laser electric field and Franck-Condon factor, whilst being common to both equations, are unknown factors. However the important factor isn't the photoionization rates but the relative change in photoionization cross section from the bulk diamond to the diamond in the potential well. To understand this the transition rates in equations \ref{eqn:TransitionBulkfinal} and \ref{eqn:Transitionelectrodefinal} are divided by the common factors mentioned and multiplied by: $1/\big(0.5 n \sqrt{\frac{\epsilon_0 \mu_0}{\hbar \omega}}\big)$ where $n$ is the refractive index of diamond and $\omega$ is the wavelength of the photoionization laser. These extra factors places the dimensions of equations \ref{eqn:TransitionBulkfinal} and \ref{eqn:Transitionelectrodefinal} into a photoionization cross section multiplied by the electric charge squared, divided by the dipole moment squared, which is a dimensionless quantity. With these changes, the two curves in figure \ref{fig:transition} can be plotted together, and the relative cross-sections can be compared.   

\section{Broadening from the electrode}

In order to create the potential well in the diamond, an electric potential must be applied to the electrode. If however, this potential is unstable, then it will cause a fluctuating shift in the discretized diamond conduction band energy levels which will present as linewidth broadening. To model this effect, the simulation can be performed with a 10~mV electrode potential and it can also be performed with and offset based on the expected noise in the signal generator: $\pm 0.001$~mV. The change in the potential at the NV (50~nm from the surface of the diamond) can be measured and the change first conduction band eigenenergy levels can also be measured. 

\begin{figure}[!ht]
    \centering
    \includegraphics[width = 0.4\textwidth]{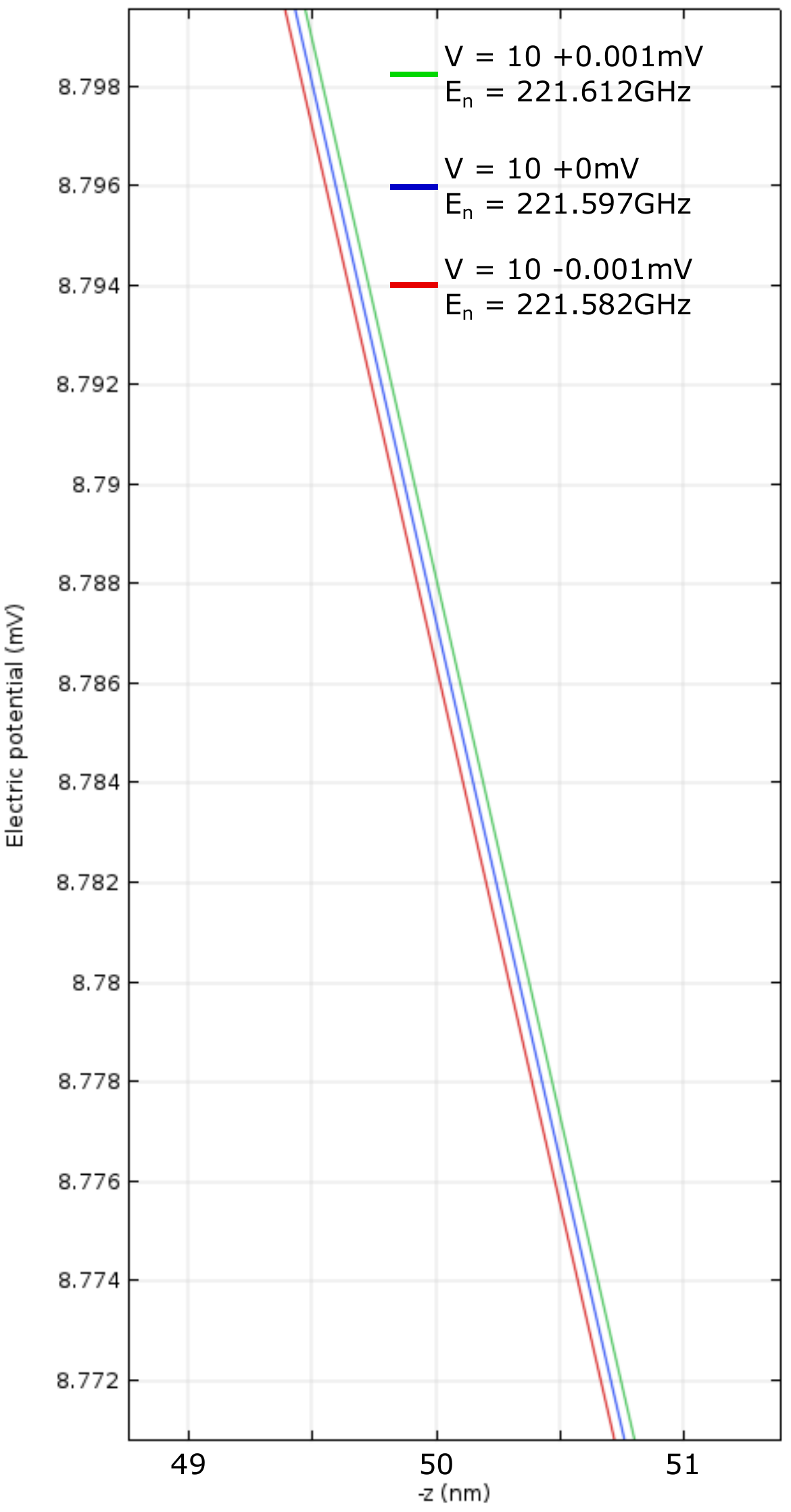}
    \caption{Plots of the electric potential as a function of the distance from the surface of the diamond where it meets the electrode. The 50~nm mark is where the NV is placed. The three curves designate the solution for the 10~mV potential as well as its offsets of $\pm 0.001$~mV. Additionally, the plot legends show the first conduction band eigenenergy levels for each potential solution. Overall the change generated from the potential noise very small.}
    \label{fig:electrodenoise}
\end{figure}

\begin{figure}[!ht]
    \centering
    \includegraphics[width = 0.45\textwidth]{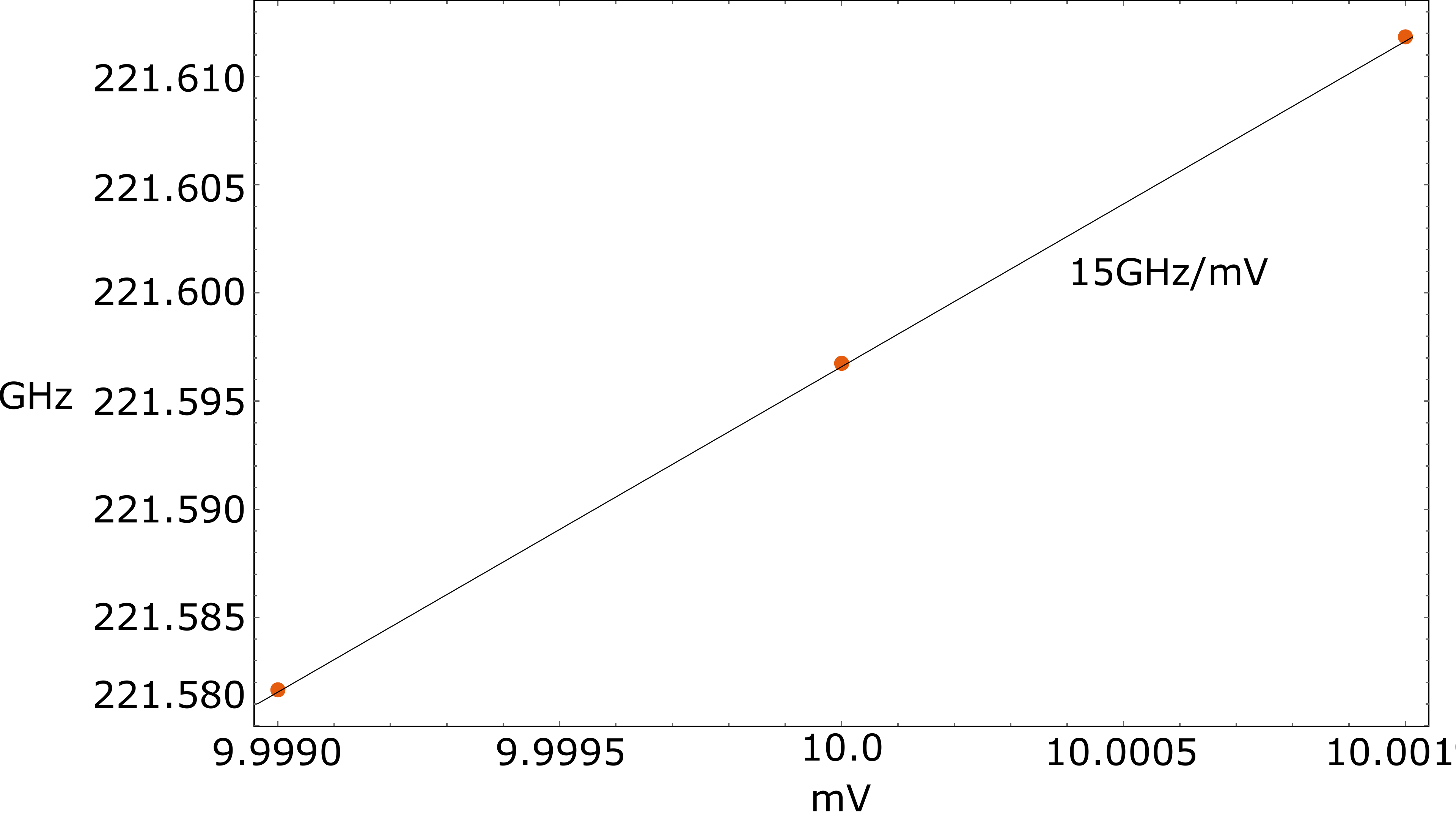}
    \caption{Plots of the first conduction band eigenenergy levels as a function of electrode potential. The change is roughly linear so a line can be drawn to connect the three data points and a slope can be calculated to be $\approx 15$~GHz/mV which is labelled on the plot.}
    \label{fig:electrodenoise2}
\end{figure}

Figure \ref{fig:electrodenoise} shows the electric potential as a function of distance from the diamond surface for all three electrode potential values. The figure also has the first conduction band eigenenergy levels labelled for each electrode potential. By taking these energy levels and plotting them the slope of the values can be calculated to be $\approx 15$~GHz/mV which is shown in figure \ref{fig:electrodenoise2}. The change in the transition energy (linewidth broadening) will then be the change in the conduction band energy as a function of potential ($\frac{\partial E_n}{\partial V} \approx 15$~GHz/mV), minus the change in the NV energy as a function of potential ($\frac{\partial E_{NV}}{\partial V}$), multiplied by the RMS uncertainty of the electrode: 

\begin{equation}\label{eqn:electrode2}
    \Gamma_{electrode} = \sigma_V\Big(\frac{\partial E_c}{\partial V}-\frac{1}{h} \frac{\partial E_{NV}}{\partial V} \Big),
\end{equation}

where $\sigma_V$ is the RMS uncertainty of the electrode which was assumed to be $\pm 0.001$~mV and the change in NV energy with potential is linear ($\frac{\partial e_{NV}}{\partial V} = -eV$) which can be calculated and converted to $-242 $~GHz/mV. Solving equation \ref{eqn:electrode2} gives the linewidth broadening mentioned in the main paper of $0.257$~GHz broadening.  

\section{Broadening from e-p scattering}

Electron-phonon (e-p) scattering is the process by which the energy of an electron is altered slightly by a phonon interaction. This will alter the transition energy of photoionization as the conduction band electron states are altered by phonon interactions, causing linewidth broadening. In bulk diamond this can be an issue even at low temperatures (4~K). We can model this broadening with Fermi's golden rule: 

\begin{equation}\label{eqn:e-p2}
    \begin{split}
    \Gamma_{ep} = \alpha\sum_n \int |\bra{n}& \psi_k (\vec{r}) \ket{1}|^2 \omega_k^3 n_B(\omega_k) \delta(\omega_n - \omega_k)d^{3}k
    \\&\alpha = \frac{\Theta^2}{2(2\pi)^2\hbar\rho c_{l}^4},
    \end{split}
\end{equation}

where $\Theta$ is the acoustic deformation potential, $\rho$ is the density of diamond and $c_l$ is the longitudinal speed of sound in diamond. The equation is a sum of all the interactions of the phonon modes $\psi_k(\vec{r})$ with the discretised energy levels of the conduction band minimum from 1 to n, integrated over the phonon k-space. The phonons have a wavelength $\omega_k$ and a temperature dependent distribution given by a Bose-Einstein distribution $n_B$ \cite{Oberg2019SpinDiamond}. The phonon modes can be calculated by understanding that e-p scattering only occurs with dilational modes as they are the only modes with a non-zero divergence. Deriving an expression for the dilational modes as: 

\begin{equation}\label{eqn:mode}
    \vec{u}(r) = \vec{\nabla}\psi(r), 
\end{equation}

the scalar potential is satisfied by the wave equation:

\begin{equation}\label{eqn:mode2}
    -\nabla^2\psi(r) = c_l^{-2} \omega^2 \psi(r),
\end{equation}

which has the following solution:

\begin{equation}\label{eqn:mode3}
    \psi(r) = A_k e^{i\vec{k}\cdot \vec{r}}.
\end{equation}

In equation \ref{eqn:mode3} the normalisation constant $A_k$ can be found by integrating all the dilational modes over the volume of the diamond:

\begin{equation}
    \begin{split}
    V = \int_V |\vec{u}|^2 d^3 r
    \\= \int_V |\vec{\nabla}\psi|^2 d^3 r
    \\= \int_V A_k^2 k^2 d^3 r
    \\A_k= k^{-1} = c_l/\omega.
    \end{split}
\end{equation}

To help solve equation \ref{eqn:e-p2} its easier to assume that confining potential is rectangular in shape rather than the Gaussian-like shapes they actually are as it allows for integrals to be solved in Cartesian coordinates whose solutions are similar to solutions to a finite square well: 

\begin{equation}\label{eqn:Gn}
    G_n(k) = |\bra{n} \psi_k (\vec{r}) \ket{1}|^2 = \Big|\int_0^{L_x}\int_0^{L_y}\int_0^{L_z} F_n \frac{c_l}{\omega_k}e^{i\vec{k}\dot\vec{r}} F_1\Big|^2,
\end{equation} 

where $L$ is the length of the confining potential in a Cartesian direction and $F_n$ is the envelope wavefunction for an electron in a Cartesian box. This assumption shouldn't change the broadening by much as long the volumes of the approximate and actual confining potentials are roughly the same. Substituting equation \ref{eqn:Gn} into equation \ref{eqn:e-p2} and writing out the integral over k gives:

\begin{equation}\label{eqn:e-p3}
    \begin{split}
    \Gamma_{ep} = \alpha \sum_n \int_0^\infty \int_0^\pi \int_0^{2\pi} k^2 
    \sin{(\theta_k)}\\ G_n(k) \omega_k^3 \eta_B(\omega_k) \delta(\omega_n-\omega_k) dk d\theta_k d\phi_k,
    \end{split}
\end{equation}

substituting $k = \omega/c_l$ and simplifying gives: 

\begin{equation}\label{eqn:e-p4}
    \begin{split}
    \Gamma_{ep} = \frac{\alpha}{c_l^3}&\sum_n \int_0^\infty \omega^5 \eta_B(\omega) \delta(\omega_n-\omega)G_n\big(\frac{\omega^2}{c_l^2}\big)\\ &\int_0^\pi \int_0^{2\pi}\sin{(\theta_k)}  d\omega^2 d\theta_k d\phi_k,
    \end{split}
\end{equation}

which when solving the integral over the Dirac delta function gives: 

\begin{equation}\label{eqn:e-p5}
    \begin{split}
    \Gamma_{ep} = \frac{\alpha}{c_l^3}&\sum_n \omega_n^5 \eta_B(\omega) G_n\big(\frac{\omega^2}{c_l^2}\big)\\ &\int_0^\pi \int_0^{2\pi}\sin{(\theta_k)}  d\omega^2 d\theta_k d\phi_k.
    \end{split}
\end{equation}

The solution to equation \ref{eqn:Gn} in Cartesian coordinates is effectively the solution to a three dimensional finite square well which can be analytically calculated to be: 

\begin{equation}\label{eqn:Gn2}
    \begin{split}
    G_n(k) = \frac{8}{L_x L_y L_z} \frac{c_l}{\omega}\\\int_0^{L_x} \sin{\Big(\frac{n_x\pi}{L_x}x\Big)}e^{ik_x x}\sin{\Big(\frac{\pi}{L_x}\Big)}dx\\\int_0^{L_y} \sin{\Big(\frac{n_y\pi}{L_y}y\Big)}e^{ik_y y}\sin{\Big(\frac{\pi}{L_y}\Big)}dy\\\int_0^{L_z} \sin{\Big(\frac{n_z\pi}{L_z}z\Big)}e^{ik_z z}\sin{\Big(\frac{\pi}{L_z}\Big)}dz,
    \end{split}
\end{equation}

where each integral in a Cartesian direction can be calculated to be: 

\begin{equation}\label{eqn:Gn3}
    \begin{split}
   \int_0^{L_m} \sin{\big(\frac{n_m\pi}{L_m}m\big)}e^{ik_m m}\sin{\big(\frac{\pi}{L_m}\big)}dm = \\\frac{2i(1+(-1)^{n_m}e^{ik_m L_m})k_m L_m^2 n_m\pi^2}{k_m^4 L_m^4 - 2k_m^2 L_m^2(1+n_m^2)\pi^2 + (-1+n_m^2)^2\pi^4}.
    \end{split}
\end{equation}

Using the solutions from equations \ref{eqn:Gn2} and \ref{eqn:Gn3} they can be substituted into equation \ref{eqn:e-p5} and solved for a given confining volume across the number of energy levels solved in the system (18). It is important to test the solution across a range of energy levels to observe when the sum of broadening values converges, 18 levels is sufficient for convergence in this case. This process was then performed multiple times over many confining potential sizes in order to understand how the broadening changes with volume.  

\begin{figure}[!ht]
    \centering
    \includegraphics[width = 0.45\textwidth]{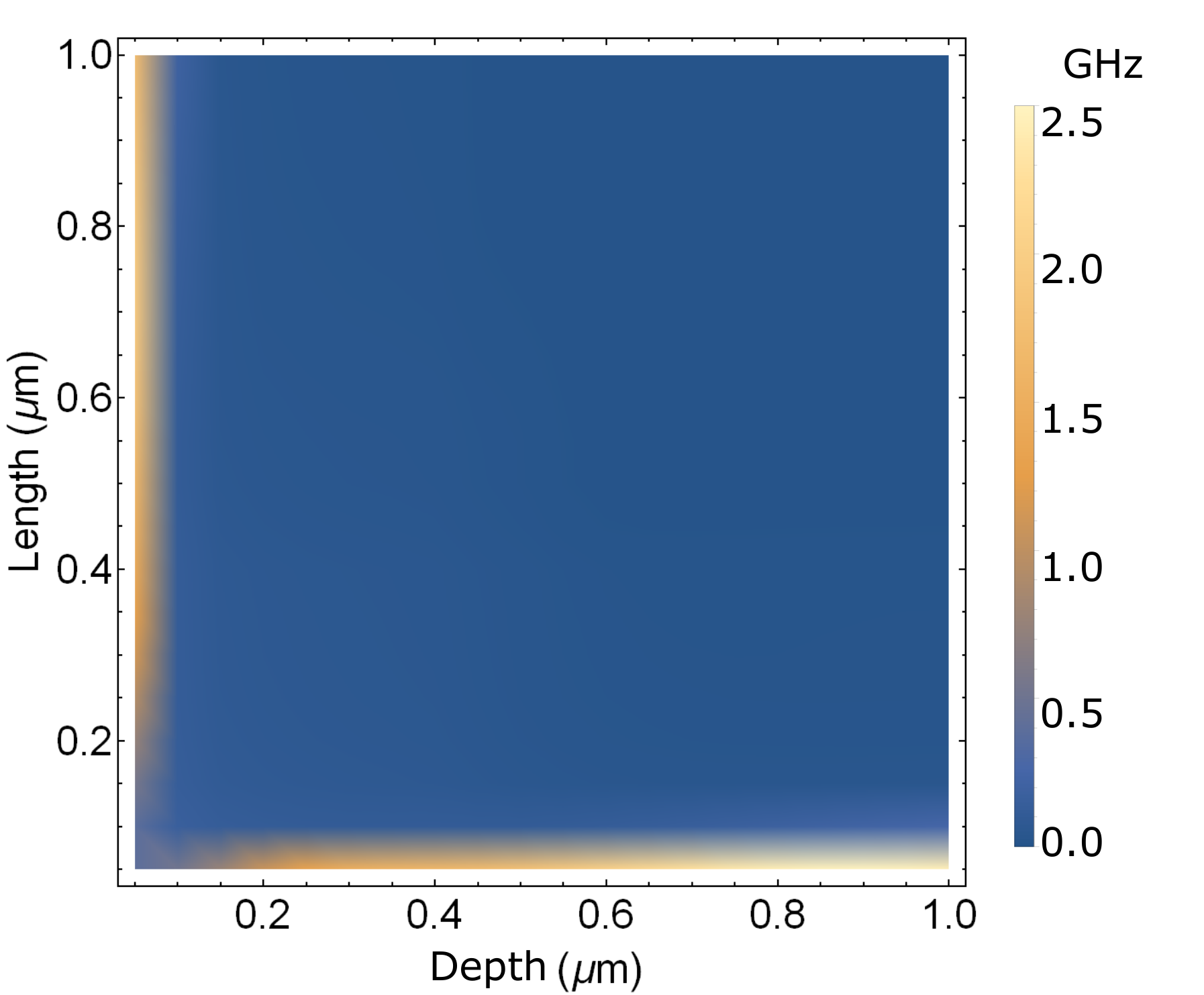}
    \caption{Plot of the e-p broadening in Hz as a function of confining potential volume. The length of the potential well is for both the x and y coordinate whereas the depth is from the z-coordinate only.}
    \label{fig:e-pBroad}
\end{figure}

In Figure \ref{fig:e-pBroad} the broadening is calculated for a variety of confining potential volumes. The x and y coordinates are changed together for the length of the confining potential and the z coordinate is changed separately for the depth of the potential well. For the purposes of this work, the broadening for a potential well that is 250~nm in length and 100~nm in depth is required. Reading off of figure \ref{fig:e-pBroad} a confining potential of roughly this size creates linewidth broadening which does not exceed 0.5~GHz.  

\section{Broadening from surface charges}

Imperfections in the diamond structure as well as its surface termination allow for surface charge traps that can hold electrons \cite{Sasama2020Charge-carrierTransistors}. Surface charge traps can be occupied briefly by electrons both from the NV as well as from defects in the bulk diamond. Charges from these surface traps can then ionize back into the bulk diamond or can hop from trap to adjacent trap. The result is a continuous fluctuation of a local electric field from the constant change in the position of the charges relative to NV. This arises as noise which can affect both the NV and conduction band energy. Calculating this noise is more difficult compared to the other sources of broadening as the density of the charge traps on the diamond surface and the rate of their motion from trap to trap isn't well known. To calculate the broadening we assume that the surface charge traps are uniformly distributed on the diamond and that the electrons occupying the traps can only move from trap to adjacent trap. In this picture, the charge motion acts like a two level fluctuator which can be modelled using Redfield theory \cite{Silchter1996PrinciplesResonance} where the linewidth is given by: 

\begin{equation}\label{eqn:surfacecharge2}
    \Gamma_{sc} = \frac{2\pi}{\hbar^2} \overline{|\delta E|^2} S(0),
\end{equation}

where $S(\omega)$ is the noise spectrum which in a two level system is assumed to be Lorentzian and $\overline{|\delta E|^2}$ is the variance of the change in energy which is given by:

\begin{equation}\label{eqn:deltaE2}
    \overline{|\delta E|^2} = \eta e^2 \int_{\Omega} |\bra{F_0}V(\vec{r},\vec{q})\ket{F_0}-V(\vec{r}_{NV}, \vec{q})|^2 d^2q,
\end{equation}

where $\eta$ is the surface density of acceptors (diamond surface trap density) and $V(\vec{r},\vec{q})$ is the potential generated by an electron at position $\vec{q}$ on the surface of the diamond away from the NV which is at position $\vec{r}$. Equation \ref{eqn:deltaE2} describes the energy variance as the change in the conduction band energy (first term in the integral) minus the change in the NV energy (second term in the integral). If the fluctuations are assumed to have a zero mean, then the rate of charge motion will follow Boltzmann statistics:

\begin{equation}\label{eqn:gamma}
    \gamma = \gamma_0 e^{-E_t/k_b T},
\end{equation}

where $E_t$ is the electron energy and the noise spectrum can be described as: 

\begin{equation}\label{eqn:specNoise}
    S(\omega) = \int_{-\infty}^{\infty} e^{-\gamma |\tau|}e^{-i\omega t} = \frac{\gamma/\pi}{\omega^2+\gamma^2},
\end{equation}

note how equation \ref{eqn:specNoise} describes the spectral noise with a Lorentzian line shape with a width given by the charge hopping rate $\gamma$. 

To get an understanding of the process and to solve equation \ref{eqn:deltaE2} for the linewidth three separate regimes are considered. The first regime is when the Fermi level is equal to the trap energy level and the trap occupation is $\approx50\%$, in this regime the electrons in the trap can hop to adjacent traps and we can model the system like a series of 2D dipoles where the dipoles can orient their direction as the electrons hop into adjacent traps in any direction. The second is when the Fermi level is below the trap energy and the trap occupation is less than $50\%$, in this regime the charges act like monopoles and can hop anywhere in the surface via other traps or through the diamond conduction band. In this regime the charges effectively appear and disappear in the traps as they have a larger freedom of movement compared to the dipole regime. The third regime is when the Fermi energy is above the trap energy level causing the traps to be mostly occupied. Charge hopping occurs in a similar mode to the monopole regime, but the effective trap density is considerably lower. This mechanism doesn't happen automatically, but it is something that can be engineered by reducing the trap density, or, adding an electron donor level in the diamond which preferentially donates electrons to the unoccupied traps. 

The first regime is the dipole regime, in this approach, the potential term in equation \ref{eqn:deltaE2} is modelled as a dipole charge: 

\begin{equation}\label{eqn:dipoleV}
    V(r) = \frac{e \vec{p}\cdot(\vec{r}-\vec{q})}{4\pi \epsilon_d |\vec{r}-\vec{q}|^3},
\end{equation}

where $\vec{p}$ is the displacement vector between neighboring traps. The idea is that an electron can move from one trap to another, creating an electron/hole pair that is effectively a dipole. The electron motion to a trap is then modelled as the dipole moment flipping sign as the electron flips the direction of the dipole. The change in conduction band energy due to a single dipole is solved individually. This solution is then integrated over all possible charge traps (the diamond surface) to find out the overall energy change. Beginning with the solution due to a single dipole:

\begin{equation}\label{eqn:Dipole}
    \bra{F_0} V(\vec{r},\vec{q}) \ket{F_0} = \frac{e}{4\pi \epsilon_d} \int F_0^2 (\vec{r}) \frac{\vec{p}\cdot(\vec{r}-\vec{q})}{4\pi \epsilon_d |\vec{r}-\vec{q}|^3} d^3 r,
\end{equation}

in this instance, $F_0$ is modelled as a Gaussian. Mathematically this approach is actually equivalent to solving for the potential of an observer point far from a dipole charge distribution, however, the roles are reversed, $F_0$ acts like the charge distribution and the observer point is $\vec{q}$. In this formalism the solution is obtained by using a multipole expansion. The key assumption in the expansion is that the observer point $\vec{q}$ is sufficiently far from the charge distribution $F_0$. This isn't entirely true for this system, so higher order terms are added in the multipole expansion to make sure it is valid. Considering the charge distribution in terms of the electric displacement:

\begin{equation}\label{eqn:D}
    \vec{\nabla} \cdot \vec{D} = 0,
\end{equation}

in equation \ref{eqn:D}, the solution is zero as the the charge density of a group of dipoles will be zero. Substituting in the displacement field with the electric field ($\vec{E}$) and polarization ($\vec{P}$) and rearranging gives:

\begin{equation}\label{eqn:D2}
    \vec{\nabla} \cdot\epsilon_d\vec{E} = -\vec{\nabla} \cdot \vec{P},
\end{equation}

the polarization can be described in terms of the wavefunction size and the displacement vector between neighboring traps ($\vec{p}$). The electric field can also be described in terms of the electric potential, giving: 

\begin{equation}\label{eqn:D3}
    \epsilon_d\nabla^2 V = -\vec{\nabla} \cdot e F_0^2 \vec{p},
\end{equation}

electrons on the diamond surface can only form dipoles on the 2D surface plane: 

\begin{equation}\label{eqn:D4}
    \nabla^2 V = \frac{-e}{\epsilon_d} \Big(p_x\frac{\partial }{\partial x} F_0^2 +  p_y\frac{\partial}{\partial y} F_0^2 \Big).
\end{equation}

Using Maxwell's law, the electric potential can be re-written in terms of the effective charge density: $\nabla^2V = \rho_{eff}/\epsilon_d$: 

\begin{equation}
    \rho_{eff} = -e\Big(p_x\frac{\partial }{\partial x} F_0^2 +  p_y\frac{\partial}{\partial y} F_0^2 \Big), 
\end{equation}

with an effective charge, an equation \ref{eqn:Dipole} be rewritten for the change in conduction band energy due to a surface trap with a multipole expansion: 

\begin{equation}\label{eqn:Dipole2}
    \begin{split}
    \bra{F_0} V(\vec{r},\vec{q}) \ket{F_0} = \frac{e}{4\pi \epsilon_d} \int \frac{\rho_{eff}}{|\vec{r}-\vec{q}|} d^3 r \\ \approx \frac{e p_0}{4\pi \epsilon_d |\vec{r}-\vec{q}|} + \sum_i \frac{e p_{1i}(r_{0i}-q_i)}{4\pi \epsilon_d |\vec{r_0}-\vec{q}|^3} +\\ \frac{1}{2}\sum_{i,j}\frac{e p_{2ij}(r_{0i}-q_i)(r_{0j}-q_j)}{4\pi \epsilon_d |\vec{r_0}-\vec{q}|^5},
    \end{split}
\end{equation}

where: 

\begin{equation}\label{eqn:peff}
    \begin{split}
    p_0 = \int& \rho_{eff} d^3 r  \\ p_{1i} = -\int \rho_{eff}&(r_{0i}-r_i) d^3r\\ p_{2ij} = \int \rho_{eff}\Big(3 (r_{0i}-r_i)&(r_{0j}-r_j) -\delta_{ij}(r_{0i}-r_i)^2\Big) d^3 r,
    \end{split}
\end{equation}

in the above equation, $r_0$ is the position of the electron in the conduction band (which is set at the origin), and the sum over i and j are for the different dimensions of the problem (x,y,z). The first term then denotes an asymmetric solution which is the monopole term, the second is the dipole and the third is the octupole term. The expansion increases with higher orders of $\vec{q}$, thus higher order terms should contribute less and less to the overall solution. Equation \ref{eqn:Dipole2} will be the solution for the conduction band electron, and the solution to the NV electron can be written from equation \ref{eqn:Dipole} where $r = r_0$ as the origin is set at the centre of the confined wavefunction where the NV is. The solutions to a single dipole charge are then integrated over the diamond surface to account for the effects of all surface charges. Additionally, the equations are also integrated over all the possible orientations of the dipole, $p_{\phi}$, where $p_x = p\cos{(p_{\phi})}$ and $p_y = p\sin{(p_{\phi})}$. The last thing to consider is the trap pair distance which will be estimated to be 1nm. This estimation comes from the density of traps, $p = \sqrt{1/\eta}$ where $\eta$ is the density of traps on the diamond surface. From this derivation, equation \ref{eqn:deltaE2} becomes: 

\begin{equation}\label{eqn:deltaE3}
    \begin{split}
    \overline{|\delta E|^2} = \frac{\eta/2 e^2}{2\pi}\int_0^{2\pi}\int_{\Omega} \frac{p_{dip}\cdot(\vec{r}-\vec{p})}{4\pi \epsilon_d|\vec{r_0}-\vec{q}|^3} -\\\frac{e p\cdot(\vec{r}-\vec{p})}{4\pi \epsilon_d|\vec{r_0}-\vec{q}|^3} + \frac{p_{hex}\cdot(\vec{r}-\vec{p})}{4\pi \epsilon_d|\vec{r_0}-\vec{q}|^7} d^2 q dp_{\phi},
    \end{split}
\end{equation}

where $\eta$ has been halved as the dipoles effectively halve the number of contributing traps. Due to the fact that the charge density is Gaussian, an even function, equation \ref{eqn:peff} shows that the first order term in the expansion will be an odd function and will integrate to zero. In fact, all odd terms in the expansion will be odd functions, so only the dipole ($p_{dip}$) and hexapole ($p_{hex}$) terms remain in the above equation which are derived from equation \ref{eqn:deltaE3}. When adding the dipole orientations in the integral, equation \ref{eqn:deltaE3} becomes:

\begin{equation}\label{eqn:deltaE4}
    \begin{split}
    \overline{|\delta E|^2} = \frac{\eta/2 e^2}{4\pi}\int_0^{2\pi}\int_{\Omega} \frac{e \big(p \cos{(p_{\phi})}+ p \sin{(p_{\phi})}\big)\cdot(\vec{r}-\vec{p})}{4\pi \epsilon_d|\vec{r_0}-\vec{q}|^3} \\-\frac{e \big(p \cos{(p_{\phi})}+ p \sin{(p_{\phi})}\big)\cdot(\vec{r}-\vec{p})}{4\pi \epsilon_d|\vec{r_0}-\vec{q}|^3} + \frac{p_{hex}\cdot(\vec{r}-\vec{p})}{4\pi \epsilon_d|\vec{r_0}-\vec{q}|^3} d^2 q dp_{\phi},
    \end{split}
\end{equation}

note how with the expansion of the dipole terms it becomes clear that the first two terms of the integral will cancel each other out, leaving just the hexapole term which has an analytic solution that will be referred to as 
$\overline{|\delta E|^2}_{d}$: 

\begin{equation}\label{eqn:deltaE5}
    \begin{split}
    \overline{|\delta E|^2}_d = \frac{15p^2e^4\eta\big(\alpha + \beta\big)}{8192\pi q_z^6 \epsilon_d^2}\\
    \alpha = 3\sigma_x^4+2\sigma_x^2 \sigma_y^2 \\
    \beta = 3\sigma_y^4 - 8(\sigma_x^2+\sigma_y^2)\sigma_z^2+&8\sigma_z^4,
    \end{split}
\end{equation}

where $\sigma_n$ is the standard deviation in a Cartesian direction for the Gaussian approximation to the wavefunction and $q_z$ is the distance from the diamond surface to the NV. Using the Lorentzian spectral noise density from equation \ref{eqn:specNoise} the linewidth broadening due to a dipole surface of charges will then become: 

\begin{equation}\label{eqn:GammaDipole}
    \Gamma_{dipole} = \frac{2\pi}{\hbar^2}\overline{|\delta E|^2}_d \frac{\gamma/\pi}{\gamma^2},
\end{equation}

the $\sigma$ values are chosen for a wavefunction confined to a volume tht is that is 250~nm in length/width and 100~nm in depth, the NV is 50~nm in depth ($q_z$) and the density of traps is $10^{18}$~m$^{-2}$. Whilst the charge hopping rate ($\gamma$) is unknown, a range of values from kHz to GHz can be tested see its effect on the overall linewidth. Even when considering extremely fast charge motion on the surface (THz in order) the broadening can be as high as $10^{15}$~Hz for a typical trap density ($10^{18}$~m$^{-2}$), as much as 6 orders of magnitude broader than the limiting requirement.  

Calculating the broadening in the monopole regime is largely the same as the dipole regime but the potential and the subsequent expansion will be for a monopole charge source. Following the logic from equation \ref{eqn:Dipole2}:

\begin{equation}\label{eqn:Monopole}
    \begin{split}
    \bra{F_0} V(\vec{r},\vec{q}) \ket{F_0} = \frac{e}{4\pi \epsilon_d} \int \frac{F_0^2}{|\vec{r}-\vec{q}|} d^3 r \\ \approx \frac{e p_0}{4\pi \epsilon_d |\vec{r}-\vec{q}|} + \sum_i \frac{e p_{1i}(r_{0i}-q_i)}{4\pi \epsilon_d |\vec{r_0}-\vec{q}|^3} +\\ \frac{1}{2}\sum_{i,j}\frac{e p_{2ij}(r_{0i}-q_i)(r_{0j}-q_j)}{4\pi \epsilon_d |\vec{r_0}-\vec{q}|^5}, 
    \end{split}
\end{equation}

where: 

\begin{equation}\label{eqn:Monopeff}
    \begin{split}
    p_0 = \int& F_0^2 d^3 r  \\ p_{1i} = -\int F_0^2&(r_{0i}-r_i) d^3r\\ p_{2ij} = \int F_0^2\Big(3 (r_{0i}-r_i)&(r_{0j}-r_j) -\delta_{ij}(r_{0i}-r_i)^2\Big) d^3 r,
    \end{split}
\end{equation}

and the octupole term can be rewritten as: 

\begin{equation}\label{eqn:MonoOct}
    \begin{split}
    p_{2ij} = \frac{1}{2}\sum_i \frac{(2\sigma_i^2 - \sum_{j\ne i}\sigma_j^2)(r_{0i}-q_i^2)}{|\vec{r_0}-\vec{q}|^5}.
    \end{split}
\end{equation}

Equations \ref{eqn:Monopole} and \ref{eqn:Monopeff} are almost exactly the same as equations \ref{eqn:Dipole2} and \ref{eqn:peff} with the exception of the substitution of $F_0^2$ for $\rho_{eff}$ to account for the dipole nature of the charge distribution in the dipole regime. This will change the non-zero terms in the expansion as $F_0^2$ is an even function whereas $\rho_{eff}$ is an odd function. This means that whilst in the dipole regime the odd terms in the expansion integrated to zero, in the monopole regime the even terms in the expansion will integrate to zero. Placing equations \ref{eqn:Monopole} and \ref{eqn:MonoOct} into equation \ref{eqn:deltaE2} gives: 

\begin{equation}\label{eqn:deltaE6}
    \begin{split}
    \overline{|\delta E|^2} = \eta e^2\big(\frac{e}{4\pi \epsilon_d}\big)\int \Big|\frac{1 }{|\vec{r_0}-\vec{q}|}-\frac{1}{|\vec{r_{NV}}-\vec{q}|} + \\ \frac{1}{2}\sum_i \frac{(2\sigma_i^2 - \sum_{j\ne i}\sigma_j^2)(r_{0i}-q_i^2)}{|\vec{r_0}-\vec{q}|^5} \Big|^2 d^2 q,
    \end{split}
\end{equation}

equation \ref{eqn:deltaE6} has an analytic solution which will now be referred to as $\overline{|\delta E|^2}_{m}$: 

\begin{equation}\label{eqn:deltaE7}
    \begin{split}
    \overline{|\delta E|^2}_m = \big(\frac{\eta e^4}{4\pi \epsilon_d}\big)^2\frac{3\pi\big(\alpha + \beta\big)}{128 q_z^4}\\
    \alpha = 3\sigma_x^4+2\sigma_x^2 \sigma_y^2 \\
    \beta = 3\sigma_y^4 - 8(\sigma_x^2+\sigma_y^2)\sigma_z^2+&8\sigma_z^4,
    \end{split}
\end{equation}

substituting equation \ref{eqn:deltaE7} into equation \ref{eqn:surfacecharge2} with the Lorentzian noise spectrum gives the linewidth broadening in the monopole regime: 

\begin{equation}\label{eqn:GammaMonopole}
    \Gamma_{monopole} = \frac{2\pi}{\hbar^2}\overline{|\delta E|^2}_m \frac{\gamma/\pi}{\gamma^2},
\end{equation}

where the parameters are chosen for the same simulation as the dipole regime. Even when considering extremely fast charge motion on the surface (THz in order) the broadening can be as high as $10^{17}$~Hz which is 2 orders higher than the dipole regime. This is likely due to the fact that the effective charge density is higher in the monopole case and that charges have a much higher freedom of movement across the diamond surface compared to the adjacent hopping in the dipole case. In both regimes the broadening is too high for resolving conduction band states individually. So a new approach is needed to reduce the effective trap density. One option is to reduce the physical number of surface traps, which is difficult to achieve, and the other is to fill the traps with a donor layer in the diamond.

When adding a donor layer of nitrogen below the NV, the surface charges and the donor layer act like the plates of a capacitor. The donor layer nitrogen's will pass their electron to the surface traps, resulting in a complete filling of the traps. The filling of the traps will mitigate any opportunity for electrons to hop within the traps, thereby reducing the AC electric field they might produce. How much trap filling that occurs will depend on the donor layer as well as the trap density. This can be described by using an effective charge density with the following capacitor equation:

\begin{equation}\label{eqn:saturatedcharge}
    \rho_s = CV = e \eta \frac{1}{e^{\frac{E_T-E_N+V(\eta,N)}{k_b T}+1}}.
\end{equation}

Where $C$ is the capacitance generated by the trap layer and the donor layer, $E_T$ is the trap energy, $E_N$ is the donor layer energy and $V(\eta,N)$ is the potential generated in eV. The exponential function is the Fermi-Dirac distribution which modulates the normal trap density $\eta$ based on the energy of the system. As long as the density of donors is high enough, the Fermi energy will be pinned to the donor layer energy and can be removed from equation \ref{eqn:saturatedcharge}. Fermi pinning is a process where the surface charge density gets high enough that the energy of the surface pins the Fermi energy at the same point. In other words, higher dopant concentrations cause charges to go to the surface but the surface charges in equilibrium force the charges to move back into the bulk, pinning the energy at a point where there is only have 50\% occupation at minimum no matter how dense the dopant concentration is compared to the trap density. Fermi pinning is in part what motivates a reduction of the trap density instead of just increasing the dopant concentration as pinning won't occur if the surface trap density is low. The only unknown in equation \ref{eqn:saturatedcharge} is the potential, $V(\eta,N)$; to solve for it equation \ref{eqn:saturatedcharge} is rearranged for the trap density and the capacitance is replaced by an equation with a capacitance per unit area for a parallel plate capacitor: $C = \epsilon_d/N$ where $N$ is the distance from the diamond surface to the nitrogen donor layer: 

\begin{equation}\label{eqn:saturatedcharge2}
    \eta = \frac{e N}{\epsilon_d V(\eta,N)} \frac{1}{e^{\frac{E_T-E_N+V(\eta,N)}{k_b T}+1}},
\end{equation}

equation \ref{eqn:saturatedcharge2} can then be solved to obtain the capacitor potential $V(\eta,N)$. The potential can be substituted into the following equation which mediates the effective trap density by the occupation of traps created by the nitrogen donor layer: 

\begin{equation}\label{eqn:saturatedcharge3}
    \begin{split}
    \eta_{eff} = \eta(1-Oc(\eta, N))\\
    Oc(\eta,N) = 0.5\Bigg(1+\frac{1}{1+e^{\frac{E_T-E_N+V(\eta,N)}{k_b T}}}\Bigg),
    \end{split}
\end{equation}

where $Oc(\eta,N)$ is the occupation of the surface traps which is a function of the initial charge density as well as the distance the donor layer is from the diamond surface. Equation \ref{eqn:saturatedcharge3} can be used in place of $\eta$ in the dipole regime linewidth equation \ref{eqn:GammaDipole}.  

\begin{figure}[!ht]
    \centering
    \includegraphics[width = 0.45\textwidth]{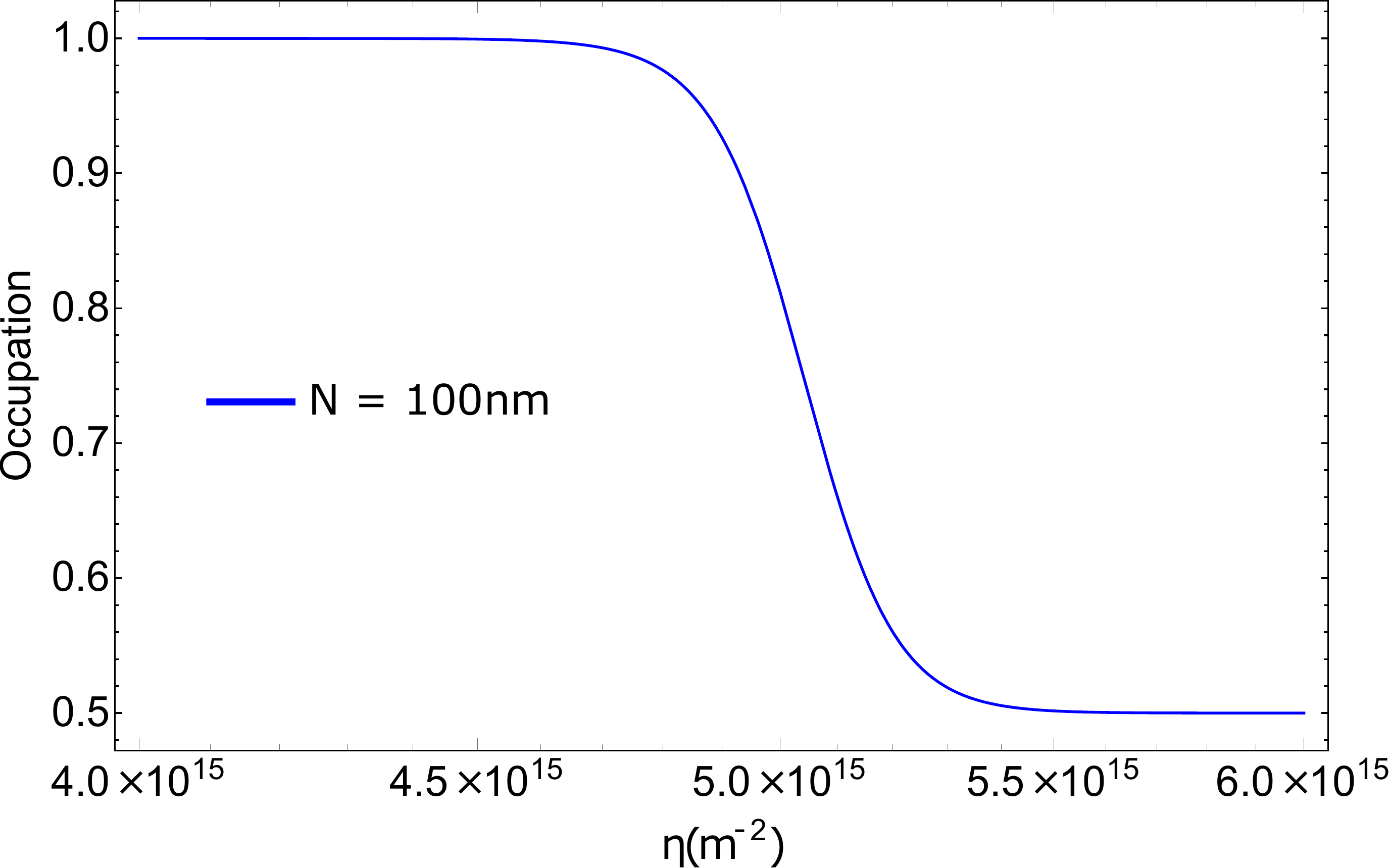}
    \caption{Plot of the trap occupation for a donor layer 100~nm below the surface of the diamond. At very low surface trap densities ($\eta < \approx 4.5\times10^{15}$~m$^{-2}$) the trap occupation is 100\% however when the density is higher, then the occupation rapidly drops to 50\%}
    \label{fig:saturation}
\end{figure}

Figure \ref{fig:saturation} shows shows the occupation as a function of the surface trap density for a donor layer that is 100~nm away from the diamond surface. From equation \ref{eqn:saturatedcharge3} it becomes clear that as the occupation increases, the effective density drops. If the traps are fully occupied, then the effective density is zero and the charge hopping will also go to zero, which occurs when the trap density is around $10^{15}$~m$^{-2}$. This means that the associated AC electric field from trap charge motion will be zero and the broadening from the surface charges will also be zero. If however, the charge density is too high, then the occupation will quickly drive towards 50\% and the linewidth broadening will become similar in value to the dipole regime. 

To ensure that there is full trap occupation there are three options: the first is to increase the capacitance by reducing the depth of the donor layer ($N$). This increases the ease of nitrogen donation. For example, when the donor layer is 50~nm instead of 100~nm then the full occupation occurs at $5.5\times10^{15}$~m$^{-2}$. There a problems with this approach, nitrogen atoms too close to the NV center decreases coherence times and the higher capacitance will cause the NV to donate electrons to the surface traps, causing unwanted ionization. The second option would be to increase the Fermi level with an external potential such as the electrode positive potential, this will increase the Fermi level thereby increasing trap occupation for the traps located around the electrode. This can be modelled by adding the electrode potential to the Fermi Dirac distribution in equations \ref{eqn:saturatedcharge}, \ref{eqn:saturatedcharge2} and \ref{eqn:saturatedcharge3}. Whilst this method will already be in effect, the potential from the electrode is relatively small (10~mV), which has a negligible effect on the occupation. If the electrode potential were to increase beyond 10~mV, then the potential well would get smaller to the point where it would be difficult to place an NV in the well volume. The third method would be to reduce the surface trap density which whilst being technically challenging, is possible and has the greatest overall effect on the occupation without affecting the NV performance. This theoretical approach motivates the final result stated in the main paper that broadening from surface traps can be completely mitigated as long as there is a nitrogen donor layer approximately 100~nm below the surface of the diamond (with the NV at 50~nm below the surface of the diamond) and the surface trap density is reduced to $\approx10^{15}$~m$^{-2}$. 

\section{Fidelity and contrast calculation}

Readout fidelity can be understood as the probability of getting the right answer in a process. Conversely the infidelity can be thought of as the probability of getting the wrong answer. The probability of getting the wrong spin state can be expressed as the sum of the probabilities for wrong processes divided by the sum of the probabilities for all processes:

\begin{equation}\label{eqn:fidelity1}
    error = \frac{p_1 + p_0}{p_0 + p_1 + p_a},
\end{equation}

In the above equation, you are trying to get a particular spin readout $p_0$, but there is a chance you could get the wrong spin state, $p_1$, during photoionization or there is a chance you could get no photoionization at all, this is because for the expected photoionization energy (2.6~eV \cite{Aslam2013Photo-inducedDetection}), there is a non-zero chance of absorption which is given by $p_a$ where the electron goes to the excited state in the NV instead of ionizing. The probability of the correct readout ($p_0$) is given by the cross section of photoionization at 2.6~eV in bulk diamond ($\sigma_{i,b}$), multiplied by a further factor, $f = 2.4$, which describes the increase in photoionization rate due to the discrete energy levels (see figure \ref{fig:transition} of the main paper). The rate of the wrong spin readout ($p_1$) can be expressed as the rate of photoionization, $\sigma_{i,b}$, multiplied by the same $f$ factor as well as an error rate given as a Lorentzian function. The Lorentzian function describes how the energy levels and linewidth broadening affects the probability of getting the correct spin state in an ionization readout process. It is a function of the difference in energy separation between the ground state NV triplet and the conduction band energy levels as well as the total linewidth broadening:  

\begin{equation}\label{eqn:Lorentz2}
    \mathcal{L}(\Delta E) = \frac{\phi/\pi}{\Delta E^2 + \phi^2},
\end{equation}

where $\phi$ is the total linewidth broadening calculated by adding all the sources of broadening listed in this paper ($0.757$~GHz). The splitting of the NV ground state triplet ($m_s=0$ and $m_s=\pm 1$ respectively) is known to be: $\Delta D \approx 2.87$~GHz \cite{Manson2018NV--N+Diamond, Doherty2013TheDiamond} and the splitting of the first two conduction band levels under the electrode is calculated to be: $\Delta C \approx 9.945$~GHz using equation \ref{eqn:schrodinger}. The $\Delta E$ term in the Lorentzian function is then the difference in energy separation between the conduction band energy levels and the ground state NV triplet energy levels ($\Delta E = \Delta C - \Delta D = 7.075$~GHz). The final term in equation \ref{eqn:fidelity1} is the absorption probability $p_a$ which can be expressed as the photoionization cross-section multiplied by the ratio of photoionization to absorption, $g = 0.4$, at the energy of photoionization (2.6~eV). This ratio can be found from the photoionization/absorption cross section data by Razinkovas et al \cite{Razinkovas2021VibrationalDiamond}. Substituting all the relevant terms into equation \ref{eqn:fidelity1}:

\begin{equation}\label{eqn:fidelityfinal}
    \begin{split}
    error = \frac{f \sigma_{i,b} \mathcal{L}(\Delta E) + g \sigma_{i,b}}{f \sigma_{i,b} + f \sigma_{i,b} \mathcal{L}(\Delta E) + g \sigma_{i,b}}\\
     = \frac{\mathcal{L}(\Delta E) + g/f}{1 + \mathcal{L}(\Delta E) + g/f},     
    \end{split}
\end{equation}

note that equation \ref{eqn:fidelityfinal} gives the error rate or the infidelity (15\%). To find the fidelity, the difference must be used (1-0.15), which is the equation used in the main paper that gives the readout fidelity of 85\%. To calculate the optical spin contrast, the same methodology is applied, but the contrast is given by the difference in probabilities for readout divided by the probability for all processes: 

\begin{equation}\label{eqn:contrastfinal}
    \begin{split}
    C = \frac{p_0 - p_1}{p_0 + p_1 + p_a}\\
     = \frac{1- \mathcal{L}(\Delta E)}{1 + \mathcal{L}(\Delta E) + g/f},     
    \end{split}
\end{equation}

solving equation F4 gives an optical spin contrast of 85\%. The calculation of fidelity and contrast both dictate the requirements of the SCC protocol. The three main factors that improve fidelity and contrast include: smaller linewidth broadening, increasing photoionization probability and increasing the energy splitting in the low lying conduction band states.

%% file: 1Main.bbl
\begin{thebibliography}{26}%
\makeatletter
\providecommand \@ifxundefined [1]{%
 \@ifx{#1\undefined}
}%
\providecommand \@ifnum [1]{%
 \ifnum #1\expandafter \@firstoftwo
 \else \expandafter \@secondoftwo
 \fi
}%
\providecommand \@ifx [1]{%
 \ifx #1\expandafter \@firstoftwo
 \else \expandafter \@secondoftwo
 \fi
}%
\providecommand \natexlab [1]{#1}%
\providecommand \enquote  [1]{``#1''}%
\providecommand \bibnamefont  [1]{#1}%
\providecommand \bibfnamefont [1]{#1}%
\providecommand \citenamefont [1]{#1}%
\providecommand \href@noop [0]{\@secondoftwo}%
\providecommand \href [0]{\begingroup \@sanitize@url \@href}%
\providecommand \@href[1]{\@@startlink{#1}\@@href}%
\providecommand \@@href[1]{\endgroup#1\@@endlink}%
\providecommand \@sanitize@url [0]{\catcode `\\12\catcode `\$12\catcode
  `\&12\catcode `\#12\catcode `\^12\catcode `\_12\catcode `\%12\relax}%
\providecommand \@@startlink[1]{}%
\providecommand \@@endlink[0]{}%
\providecommand \url  [0]{\begingroup\@sanitize@url \@url }%
\providecommand \@url [1]{\endgroup\@href {#1}{\urlprefix }}%
\providecommand \urlprefix  [0]{URL }%
\providecommand \Eprint [0]{\href }%
\providecommand \doibase [0]{https://doi.org/}%
\providecommand \selectlanguage [0]{\@gobble}%
\providecommand \bibinfo  [0]{\@secondoftwo}%
\providecommand \bibfield  [0]{\@secondoftwo}%
\providecommand \translation [1]{[#1]}%
\providecommand \BibitemOpen [0]{}%
\providecommand \bibitemStop [0]{}%
\providecommand \bibitemNoStop [0]{.\EOS\space}%
\providecommand \EOS [0]{\spacefactor3000\relax}%
\providecommand \BibitemShut  [1]{\csname bibitem#1\endcsname}%
\let\auto@bib@innerbib\@empty
\bibitem [{\citenamefont {Barry}\ \emph {et~al.}(2016)\citenamefont {Barry},
  \citenamefont {Turner}, \citenamefont {Schloss}, \citenamefont {Glenn},
  \citenamefont {Song}, \citenamefont {Lukin}, \citenamefont {Park},\ and\
  \citenamefont {Walsworth}}]{Barry2016b}%
  \BibitemOpen
  \bibfield  {author} {\bibinfo {author} {\bibfnamefont {J.~F.}\ \bibnamefont
  {Barry}}, \bibinfo {author} {\bibfnamefont {M.~J.}\ \bibnamefont {Turner}},
  \bibinfo {author} {\bibfnamefont {J.~M.}\ \bibnamefont {Schloss}}, \bibinfo
  {author} {\bibfnamefont {D.~R.}\ \bibnamefont {Glenn}}, \bibinfo {author}
  {\bibfnamefont {Y.}~\bibnamefont {Song}}, \bibinfo {author} {\bibfnamefont
  {M.~D.}\ \bibnamefont {Lukin}}, \bibinfo {author} {\bibfnamefont
  {H.}~\bibnamefont {Park}},\ and\ \bibinfo {author} {\bibfnamefont {R.~L.}\
  \bibnamefont {Walsworth}},\ }\bibfield  {title} {\bibinfo {title} {{Optical
  magnetic detection of single-neuron action potentials using quantum defects
  in diamond}},\ }\href {https://doi.org/10.1073/pnas.1601513113} {\bibfield
  {journal} {\bibinfo  {journal} {Proceedings of the National Academy of
  Sciences}\ }\textbf {\bibinfo {volume} {113}},\ \bibinfo {pages} {14133}
  (\bibinfo {year} {2016})}\BibitemShut {NoStop}%
\bibitem [{\citenamefont {Kucsko}\ \emph {et~al.}(2013)\citenamefont {Kucsko},
  \citenamefont {Maurer}, \citenamefont {Yao}, \citenamefont {Kubo},
  \citenamefont {Noh}, \citenamefont {Lo}, \citenamefont {Park},\ and\
  \citenamefont {Lukin}}]{Kucsko2013}%
  \BibitemOpen
  \bibfield  {author} {\bibinfo {author} {\bibfnamefont {G.}~\bibnamefont
  {Kucsko}}, \bibinfo {author} {\bibfnamefont {P.~C.}\ \bibnamefont {Maurer}},
  \bibinfo {author} {\bibfnamefont {N.~Y.}\ \bibnamefont {Yao}}, \bibinfo
  {author} {\bibfnamefont {M.}~\bibnamefont {Kubo}}, \bibinfo {author}
  {\bibfnamefont {H.~J.}\ \bibnamefont {Noh}}, \bibinfo {author} {\bibfnamefont
  {P.~K.}\ \bibnamefont {Lo}}, \bibinfo {author} {\bibfnamefont
  {H.}~\bibnamefont {Park}},\ and\ \bibinfo {author} {\bibfnamefont {M.~D.}\
  \bibnamefont {Lukin}},\ }\bibfield  {title} {\bibinfo {title}
  {{Nanometre-scale thermometry in a living cell}},\ }\href
  {https://doi.org/10.1038/nature12373} {\bibfield  {journal} {\bibinfo
  {journal} {Nature}\ }\textbf {\bibinfo {volume} {500}},\ \bibinfo {pages}
  {54} (\bibinfo {year} {2013})}\BibitemShut {NoStop}%
\bibitem [{\citenamefont {Le~Sage}\ \emph {et~al.}(2013)\citenamefont
  {Le~Sage}, \citenamefont {Arai}, \citenamefont {Glenn}, \citenamefont
  {Devience}, \citenamefont {Pham}, \citenamefont {Rahn-Lee}, \citenamefont
  {Lukin}, \citenamefont {Yacoby}, \citenamefont {Komeili},\ and\ \citenamefont
  {Walsworth}}]{LeSage2013OpticalCells}%
  \BibitemOpen
  \bibfield  {author} {\bibinfo {author} {\bibfnamefont {D.}~\bibnamefont
  {Le~Sage}}, \bibinfo {author} {\bibfnamefont {K.}~\bibnamefont {Arai}},
  \bibinfo {author} {\bibfnamefont {D.~R.}\ \bibnamefont {Glenn}}, \bibinfo
  {author} {\bibfnamefont {S.~J.}\ \bibnamefont {Devience}}, \bibinfo {author}
  {\bibfnamefont {L.~M.}\ \bibnamefont {Pham}}, \bibinfo {author}
  {\bibfnamefont {L.}~\bibnamefont {Rahn-Lee}}, \bibinfo {author}
  {\bibfnamefont {M.~D.}\ \bibnamefont {Lukin}}, \bibinfo {author}
  {\bibfnamefont {A.}~\bibnamefont {Yacoby}}, \bibinfo {author} {\bibfnamefont
  {A.}~\bibnamefont {Komeili}},\ and\ \bibinfo {author} {\bibfnamefont {R.~L.}\
  \bibnamefont {Walsworth}},\ }\bibfield  {title} {\bibinfo {title} {{Optical
  magnetic imaging of living cells}},\ }\href
  {https://doi.org/10.1038/nature12072} {\bibfield  {journal} {\bibinfo
  {journal} {Nature}\ }\textbf {\bibinfo {volume} {496}},\ \bibinfo {pages}
  {486} (\bibinfo {year} {2013})}\BibitemShut {NoStop}%
\bibitem [{\citenamefont {Barson}\ \emph {et~al.}(2021)\citenamefont {Barson},
  \citenamefont {Oberg}, \citenamefont {McGuinness}, \citenamefont {Denisenko},
  \citenamefont {Manson}, \citenamefont {Wrachtrup},\ and\ \citenamefont
  {Doherty}}]{Barson2021NanoscaleSpin}%
  \BibitemOpen
  \bibfield  {author} {\bibinfo {author} {\bibfnamefont {M.~S.}\ \bibnamefont
  {Barson}}, \bibinfo {author} {\bibfnamefont {L.~M.}\ \bibnamefont {Oberg}},
  \bibinfo {author} {\bibfnamefont {L.~P.}\ \bibnamefont {McGuinness}},
  \bibinfo {author} {\bibfnamefont {A.}~\bibnamefont {Denisenko}}, \bibinfo
  {author} {\bibfnamefont {N.~B.}\ \bibnamefont {Manson}}, \bibinfo {author}
  {\bibfnamefont {J.}~\bibnamefont {Wrachtrup}},\ and\ \bibinfo {author}
  {\bibfnamefont {M.~W.}\ \bibnamefont {Doherty}},\ }\bibfield  {title}
  {\bibinfo {title} {{Nanoscale Vector Electric Field Imaging Using a Single
  Electron Spin}},\ }\href {https://doi.org/10.1021/acs.nanolett.1c00082}
  {\bibfield  {journal} {\bibinfo  {journal} {Nano Letters}\ }\textbf {\bibinfo
  {volume} {21}},\ \bibinfo {pages} {2962} (\bibinfo {year}
  {2021})}\BibitemShut {NoStop}%
\bibitem [{\citenamefont {Maurer}\ \emph {et~al.}(2012)\citenamefont {Maurer},
  \citenamefont {Kucsko}, \citenamefont {Latta}, \citenamefont {Jiang},
  \citenamefont {Yao}, \citenamefont {Bennett}, \citenamefont {Pastawski},
  \citenamefont {Hunger}, \citenamefont {Chisholm}, \citenamefont {Markham},
  \citenamefont {Twitchen}, \citenamefont {Ignacio~Cirac},\ and\ \citenamefont
  {Lukin}}]{Maurer2012Room-temperatureSecond}%
  \BibitemOpen
  \bibfield  {author} {\bibinfo {author} {\bibfnamefont {P.~C.}\ \bibnamefont
  {Maurer}}, \bibinfo {author} {\bibfnamefont {G.}~\bibnamefont {Kucsko}},
  \bibinfo {author} {\bibfnamefont {C.}~\bibnamefont {Latta}}, \bibinfo
  {author} {\bibfnamefont {L.}~\bibnamefont {Jiang}}, \bibinfo {author}
  {\bibfnamefont {N.~Y.}\ \bibnamefont {Yao}}, \bibinfo {author} {\bibfnamefont
  {S.~D.}\ \bibnamefont {Bennett}}, \bibinfo {author} {\bibfnamefont
  {F.}~\bibnamefont {Pastawski}}, \bibinfo {author} {\bibfnamefont
  {D.}~\bibnamefont {Hunger}}, \bibinfo {author} {\bibfnamefont
  {N.}~\bibnamefont {Chisholm}}, \bibinfo {author} {\bibfnamefont
  {M.}~\bibnamefont {Markham}}, \bibinfo {author} {\bibfnamefont {D.~J.}\
  \bibnamefont {Twitchen}}, \bibinfo {author} {\bibfnamefont {J.}~\bibnamefont
  {Ignacio~Cirac}},\ and\ \bibinfo {author} {\bibfnamefont {M.~D.}\
  \bibnamefont {Lukin}},\ }\bibfield  {title} {\bibinfo {title}
  {{Room-temperature quantum bit memory exceeding one second}},\ }\href
  {https://doi.org/10.1126/science.1220513} {\bibfield  {journal} {\bibinfo
  {journal} {Optics InfoBase Conference Papers}\ }\textbf {\bibinfo {volume}
  {336}},\ \bibinfo {pages} {1283} (\bibinfo {year} {2012})}\BibitemShut
  {NoStop}%
\bibitem [{\citenamefont {Wu}\ \emph {et~al.}(2019)\citenamefont {Wu},
  \citenamefont {Wang}, \citenamefont {Qin}, \citenamefont {Rong},\ and\
  \citenamefont {Du}}]{Wu2019AConditions}%
  \BibitemOpen
  \bibfield  {author} {\bibinfo {author} {\bibfnamefont {Y.}~\bibnamefont
  {Wu}}, \bibinfo {author} {\bibfnamefont {Y.}~\bibnamefont {Wang}}, \bibinfo
  {author} {\bibfnamefont {X.}~\bibnamefont {Qin}}, \bibinfo {author}
  {\bibfnamefont {X.}~\bibnamefont {Rong}},\ and\ \bibinfo {author}
  {\bibfnamefont {J.}~\bibnamefont {Du}},\ }\bibfield  {title} {\bibinfo
  {title} {{A programmable two-qubit solid-state quantum processor under
  ambient conditions}},\ }\bibfield  {journal} {\bibinfo  {journal} {npj
  Quantum Information}\ }\textbf {\bibinfo {volume} {5}},\ \href
  {https://doi.org/10.1038/s41534-019-0129-z} {10.1038/s41534-019-0129-z}
  (\bibinfo {year} {2019})\BibitemShut {NoStop}%
\bibitem [{\citenamefont {Nizovtsev}\ \emph {et~al.}(2005)\citenamefont
  {Nizovtsev}, \citenamefont {Kilin.}, \citenamefont {Jelezko}, \citenamefont
  {Gaebal}, \citenamefont {Popa}, \citenamefont {Gruber},\ and\ \citenamefont
  {Wrachtrup}}]{Nizovtsev2005ASpins}%
  \BibitemOpen
  \bibfield  {author} {\bibinfo {author} {\bibfnamefont {A.~P.}\ \bibnamefont
  {Nizovtsev}}, \bibinfo {author} {\bibfnamefont {S.~Y.}\ \bibnamefont
  {Kilin.}}, \bibinfo {author} {\bibfnamefont {F.}~\bibnamefont {Jelezko}},
  \bibinfo {author} {\bibfnamefont {T.}~\bibnamefont {Gaebal}}, \bibinfo
  {author} {\bibfnamefont {I.}~\bibnamefont {Popa}}, \bibinfo {author}
  {\bibfnamefont {A.}~\bibnamefont {Gruber}},\ and\ \bibinfo {author}
  {\bibfnamefont {J.}~\bibnamefont {Wrachtrup}},\ }\bibfield  {title} {\bibinfo
  {title} {{A quantum computer based on NV centers in diamond: Optically
  detected nutations of single electron and nuclear spins}},\ }\href
  {https://doi.org/10.1134/1.2034610} {\bibfield  {journal} {\bibinfo
  {journal} {Optics and Spectroscopy (English translation of Optika i
  Spektroskopiya)}\ }\textbf {\bibinfo {volume} {99}},\ \bibinfo {pages} {233}
  (\bibinfo {year} {2005})}\BibitemShut {NoStop}%
\bibitem [{\citenamefont {Balasubramanian}\ \emph {et~al.}(2009)\citenamefont
  {Balasubramanian}, \citenamefont {Neumann}, \citenamefont {Twitchen},
  \citenamefont {Markham}, \citenamefont {Kolesov}, \citenamefont {Mizuochi},
  \citenamefont {Isoya}, \citenamefont {Achard}, \citenamefont {Beck},
  \citenamefont {Tissler}, \citenamefont {Jacques}, \citenamefont {Hemmer},
  \citenamefont {Jelezko},\ and\ \citenamefont
  {Wrachtrup}}]{Balasubramanian2009}%
  \BibitemOpen
  \bibfield  {author} {\bibinfo {author} {\bibfnamefont {G.}~\bibnamefont
  {Balasubramanian}}, \bibinfo {author} {\bibfnamefont {P.}~\bibnamefont
  {Neumann}}, \bibinfo {author} {\bibfnamefont {D.}~\bibnamefont {Twitchen}},
  \bibinfo {author} {\bibfnamefont {M.}~\bibnamefont {Markham}}, \bibinfo
  {author} {\bibfnamefont {R.}~\bibnamefont {Kolesov}}, \bibinfo {author}
  {\bibfnamefont {N.}~\bibnamefont {Mizuochi}}, \bibinfo {author}
  {\bibfnamefont {J.}~\bibnamefont {Isoya}}, \bibinfo {author} {\bibfnamefont
  {J.}~\bibnamefont {Achard}}, \bibinfo {author} {\bibfnamefont
  {J.}~\bibnamefont {Beck}}, \bibinfo {author} {\bibfnamefont {J.}~\bibnamefont
  {Tissler}}, \bibinfo {author} {\bibfnamefont {V.}~\bibnamefont {Jacques}},
  \bibinfo {author} {\bibfnamefont {P.~R.}\ \bibnamefont {Hemmer}}, \bibinfo
  {author} {\bibfnamefont {F.}~\bibnamefont {Jelezko}},\ and\ \bibinfo {author}
  {\bibfnamefont {J.}~\bibnamefont {Wrachtrup}},\ }\bibfield  {title} {\bibinfo
  {title} {{Ultralong spin coherence time in isotopically engineered
  diamond}},\ }\href {https://doi.org/10.1038/nmat2420} {\bibfield  {journal}
  {\bibinfo  {journal} {Nature Materials}\ }\textbf {\bibinfo {volume} {8}},\
  \bibinfo {pages} {383} (\bibinfo {year} {2009})}\BibitemShut {NoStop}%
\bibitem [{\citenamefont {Doherty}\ \emph {et~al.}(2013)\citenamefont
  {Doherty}, \citenamefont {Manson}, \citenamefont {Delaney}, \citenamefont
  {Jelezko}, \citenamefont {Wrachtrup},\ and\ \citenamefont
  {Hollenberg}}]{Doherty2013TheDiamond}%
  \BibitemOpen
  \bibfield  {author} {\bibinfo {author} {\bibfnamefont {M.~W.}\ \bibnamefont
  {Doherty}}, \bibinfo {author} {\bibfnamefont {N.~B.}\ \bibnamefont {Manson}},
  \bibinfo {author} {\bibfnamefont {P.}~\bibnamefont {Delaney}}, \bibinfo
  {author} {\bibfnamefont {F.}~\bibnamefont {Jelezko}}, \bibinfo {author}
  {\bibfnamefont {J.}~\bibnamefont {Wrachtrup}},\ and\ \bibinfo {author}
  {\bibfnamefont {L.~C.}\ \bibnamefont {Hollenberg}},\ }\bibfield  {title}
  {\bibinfo {title} {{The nitrogen-vacancy colour centre in diamond}},\ }\href
  {https://doi.org/10.1016/j.physrep.2013.02.001} {\bibfield  {journal}
  {\bibinfo  {journal} {Physics Reports}\ }\textbf {\bibinfo {volume} {528}},\
  \bibinfo {pages} {1} (\bibinfo {year} {2013})}\BibitemShut {NoStop}%
\bibitem [{\citenamefont {Babinec}\ \emph {et~al.}(2010)\citenamefont
  {Babinec}, \citenamefont {Hausmann}, \citenamefont {Khan}, \citenamefont
  {Zhang}, \citenamefont {Maze}, \citenamefont {Hemmer},\ and\ \citenamefont
  {Lon{\v{c}}ar}}]{Babinec2010a}%
  \BibitemOpen
  \bibfield  {author} {\bibinfo {author} {\bibfnamefont {T.~M.}\ \bibnamefont
  {Babinec}}, \bibinfo {author} {\bibfnamefont {B.~J.}\ \bibnamefont
  {Hausmann}}, \bibinfo {author} {\bibfnamefont {M.}~\bibnamefont {Khan}},
  \bibinfo {author} {\bibfnamefont {Y.}~\bibnamefont {Zhang}}, \bibinfo
  {author} {\bibfnamefont {J.~R.}\ \bibnamefont {Maze}}, \bibinfo {author}
  {\bibfnamefont {P.~R.}\ \bibnamefont {Hemmer}},\ and\ \bibinfo {author}
  {\bibfnamefont {M.}~\bibnamefont {Lon{\v{c}}ar}},\ }\bibfield  {title}
  {\bibinfo {title} {{A diamond nanowire single-photon source}},\ }\href
  {https://doi.org/10.1038/nnano.2010.6} {\bibfield  {journal} {\bibinfo
  {journal} {Nature Nanotechnology}\ }\textbf {\bibinfo {volume} {5}},\
  \bibinfo {pages} {195} (\bibinfo {year} {2010})}\BibitemShut {NoStop}%
\bibitem [{\citenamefont {Jamali}\ \emph {et~al.}(2014)\citenamefont {Jamali},
  \citenamefont {Gerhardt}, \citenamefont {Rezai}, \citenamefont {Frenner},
  \citenamefont {Fedder},\ and\ \citenamefont {Wrachtrup}}]{Jamali2014a}%
  \BibitemOpen
  \bibfield  {author} {\bibinfo {author} {\bibfnamefont {M.}~\bibnamefont
  {Jamali}}, \bibinfo {author} {\bibfnamefont {I.}~\bibnamefont {Gerhardt}},
  \bibinfo {author} {\bibfnamefont {M.}~\bibnamefont {Rezai}}, \bibinfo
  {author} {\bibfnamefont {K.}~\bibnamefont {Frenner}}, \bibinfo {author}
  {\bibfnamefont {H.}~\bibnamefont {Fedder}},\ and\ \bibinfo {author}
  {\bibfnamefont {J.}~\bibnamefont {Wrachtrup}},\ }\bibfield  {title} {\bibinfo
  {title} {{Microscopic diamond solid-immersion-lenses fabricated around single
  defect centers by focused ion beam milling}},\ }\bibfield  {journal}
  {\bibinfo  {journal} {Review of Scientific Instruments}\ }\textbf {\bibinfo
  {volume} {85}},\ \href {https://doi.org/10.1063/1.4902818}
  {10.1063/1.4902818} (\bibinfo {year} {2014})\BibitemShut {NoStop}%
\bibitem [{\citenamefont {Wan}\ \emph {et~al.}(2018)\citenamefont {Wan},
  \citenamefont {Shields}, \citenamefont {Kim}, \citenamefont {Mouradian},
  \citenamefont {Lienhard}, \citenamefont {Walsh}, \citenamefont {Bakhru},
  \citenamefont {Schr{\"{o}}der},\ and\ \citenamefont {Englund}}]{Wan2018}%
  \BibitemOpen
  \bibfield  {author} {\bibinfo {author} {\bibfnamefont {N.~H.}\ \bibnamefont
  {Wan}}, \bibinfo {author} {\bibfnamefont {B.~J.}\ \bibnamefont {Shields}},
  \bibinfo {author} {\bibfnamefont {D.}~\bibnamefont {Kim}}, \bibinfo {author}
  {\bibfnamefont {S.}~\bibnamefont {Mouradian}}, \bibinfo {author}
  {\bibfnamefont {B.}~\bibnamefont {Lienhard}}, \bibinfo {author}
  {\bibfnamefont {M.}~\bibnamefont {Walsh}}, \bibinfo {author} {\bibfnamefont
  {H.}~\bibnamefont {Bakhru}}, \bibinfo {author} {\bibfnamefont
  {T.}~\bibnamefont {Schr{\"{o}}der}},\ and\ \bibinfo {author} {\bibfnamefont
  {D.}~\bibnamefont {Englund}},\ }\bibfield  {title} {\bibinfo {title}
  {{Efficient Extraction of Light from a Nitrogen-Vacancy Center in a Diamond
  Parabolic Reflector}},\ }\href {https://doi.org/10.1021/acs.nanolett.7b04684}
  {\bibfield  {journal} {\bibinfo  {journal} {Nano Letters}\ }\textbf {\bibinfo
  {volume} {18}},\ \bibinfo {pages} {2787} (\bibinfo {year}
  {2018})}\BibitemShut {NoStop}%
\bibitem [{\citenamefont {Aslam}\ \emph {et~al.}(2013)\citenamefont {Aslam},
  \citenamefont {Waldherr}, \citenamefont {Neumann}, \citenamefont {Jelezko},\
  and\ \citenamefont {Wrachtrup}}]{Aslam2013Photo-inducedDetection}%
  \BibitemOpen
  \bibfield  {author} {\bibinfo {author} {\bibfnamefont {N.}~\bibnamefont
  {Aslam}}, \bibinfo {author} {\bibfnamefont {G.}~\bibnamefont {Waldherr}},
  \bibinfo {author} {\bibfnamefont {P.}~\bibnamefont {Neumann}}, \bibinfo
  {author} {\bibfnamefont {F.}~\bibnamefont {Jelezko}},\ and\ \bibinfo {author}
  {\bibfnamefont {J.}~\bibnamefont {Wrachtrup}},\ }\bibfield  {title} {\bibinfo
  {title} {{Photo-induced ionization dynamics of the nitrogen vacancy defect in
  diamond investigated by single-shot charge state detection}},\ }\href
  {https://doi.org/10.1088/1367-2630/15/1/013064} {\bibfield  {journal}
  {\bibinfo  {journal} {New Journal of Physics}\ }\textbf {\bibinfo {volume}
  {15}},\ \bibinfo {pages} {0} (\bibinfo {year} {2013})}\BibitemShut {NoStop}%
\bibitem [{\citenamefont {Jaskula}\ \emph {et~al.}(2019)\citenamefont
  {Jaskula}, \citenamefont {Shields}, \citenamefont {Bauch}, \citenamefont
  {Lukin}, \citenamefont {Trifonov},\ and\ \citenamefont
  {Walsworth}}]{Jaskula2019ImprovedConversion}%
  \BibitemOpen
  \bibfield  {author} {\bibinfo {author} {\bibfnamefont {J.~C.}\ \bibnamefont
  {Jaskula}}, \bibinfo {author} {\bibfnamefont {B.~J.}\ \bibnamefont
  {Shields}}, \bibinfo {author} {\bibfnamefont {E.}~\bibnamefont {Bauch}},
  \bibinfo {author} {\bibfnamefont {M.~D.}\ \bibnamefont {Lukin}}, \bibinfo
  {author} {\bibfnamefont {A.~S.}\ \bibnamefont {Trifonov}},\ and\ \bibinfo
  {author} {\bibfnamefont {R.~L.}\ \bibnamefont {Walsworth}},\ }\bibfield
  {title} {\bibinfo {title} {{Improved Quantum Sensing with a Single
  Solid-State Spin via Spin-to-Charge Conversion}},\ }\href
  {https://doi.org/10.1103/PhysRevApplied.11.064003} {\bibfield  {journal}
  {\bibinfo  {journal} {Physical Review Applied}\ }\textbf {\bibinfo {volume}
  {11}},\ \bibinfo {pages} {1} (\bibinfo {year} {2019})}\BibitemShut {NoStop}%
\bibitem [{\citenamefont {Siyushev}\ \emph {et~al.}(2019)\citenamefont
  {Siyushev}, \citenamefont {Nesladek}, \citenamefont {Bourgeois},
  \citenamefont {Gulka}, \citenamefont {Hruby}, \citenamefont {Yamamoto},
  \citenamefont {Trupke}, \citenamefont {Teraji}, \citenamefont {Isoya},\ and\
  \citenamefont {Jelezko}}]{Siyushev2019PhotoelectricalDiamond}%
  \BibitemOpen
  \bibfield  {author} {\bibinfo {author} {\bibfnamefont {P.}~\bibnamefont
  {Siyushev}}, \bibinfo {author} {\bibfnamefont {M.}~\bibnamefont {Nesladek}},
  \bibinfo {author} {\bibfnamefont {E.}~\bibnamefont {Bourgeois}}, \bibinfo
  {author} {\bibfnamefont {M.}~\bibnamefont {Gulka}}, \bibinfo {author}
  {\bibfnamefont {J.}~\bibnamefont {Hruby}}, \bibinfo {author} {\bibfnamefont
  {T.}~\bibnamefont {Yamamoto}}, \bibinfo {author} {\bibfnamefont
  {M.}~\bibnamefont {Trupke}}, \bibinfo {author} {\bibfnamefont
  {T.}~\bibnamefont {Teraji}}, \bibinfo {author} {\bibfnamefont
  {J.}~\bibnamefont {Isoya}},\ and\ \bibinfo {author} {\bibfnamefont
  {F.}~\bibnamefont {Jelezko}},\ }\bibfield  {title} {\bibinfo {title}
  {{Photoelectrical imaging and coherent spin-state readout of single
  nitrogen-vacancy centers in diamond}},\ }\href
  {https://doi.org/10.1126/science.aav2789} {\bibfield  {journal} {\bibinfo
  {journal} {Science}\ }\textbf {\bibinfo {volume} {363}},\ \bibinfo {pages}
  {728} (\bibinfo {year} {2019})}\BibitemShut {NoStop}%
\bibitem [{\citenamefont {Gulka}\ \emph {et~al.}(2021)\citenamefont {Gulka},
  \citenamefont {Wirtitsch}, \citenamefont {Iv{\'{a}}dy}, \citenamefont
  {Vodnik}, \citenamefont {Hruby}, \citenamefont {Magchiels}, \citenamefont
  {Bourgeois}, \citenamefont {Gali}, \citenamefont {Trupke},\ and\
  \citenamefont {Nesladek}}]{Gulka2021Room-temperatureSpins}%
  \BibitemOpen
  \bibfield  {author} {\bibinfo {author} {\bibfnamefont {M.}~\bibnamefont
  {Gulka}}, \bibinfo {author} {\bibfnamefont {D.}~\bibnamefont {Wirtitsch}},
  \bibinfo {author} {\bibfnamefont {V.}~\bibnamefont {Iv{\'{a}}dy}}, \bibinfo
  {author} {\bibfnamefont {J.}~\bibnamefont {Vodnik}}, \bibinfo {author}
  {\bibfnamefont {J.}~\bibnamefont {Hruby}}, \bibinfo {author} {\bibfnamefont
  {G.}~\bibnamefont {Magchiels}}, \bibinfo {author} {\bibfnamefont
  {E.}~\bibnamefont {Bourgeois}}, \bibinfo {author} {\bibfnamefont
  {A.}~\bibnamefont {Gali}}, \bibinfo {author} {\bibfnamefont {M.}~\bibnamefont
  {Trupke}},\ and\ \bibinfo {author} {\bibfnamefont {M.}~\bibnamefont
  {Nesladek}},\ }\bibfield  {title} {\bibinfo {title} {{Room-temperature
  control and electrical readout of individual nitrogen-vacancy nuclear
  spins}},\ }\href {https://doi.org/10.1038/s41467-021-24494-x} {\bibfield
  {journal} {\bibinfo  {journal} {Nature Communications}\ }\textbf {\bibinfo
  {volume} {12}},\ \bibinfo {pages} {1} (\bibinfo {year} {2021})}\BibitemShut
  {NoStop}%
\bibitem [{\citenamefont {Zhang}\ \emph {et~al.}(2021)\citenamefont {Zhang},
  \citenamefont {Guo}, \citenamefont {Ji}, \citenamefont {Wang}, \citenamefont
  {Yin}, \citenamefont {Kong}, \citenamefont {Lin}, \citenamefont {Yin},
  \citenamefont {Shi}, \citenamefont {Wang},\ and\ \citenamefont
  {Du}}]{Zhang2021High-fidelityConversion}%
  \BibitemOpen
  \bibfield  {author} {\bibinfo {author} {\bibfnamefont {Q.}~\bibnamefont
  {Zhang}}, \bibinfo {author} {\bibfnamefont {Y.}~\bibnamefont {Guo}}, \bibinfo
  {author} {\bibfnamefont {W.}~\bibnamefont {Ji}}, \bibinfo {author}
  {\bibfnamefont {M.}~\bibnamefont {Wang}}, \bibinfo {author} {\bibfnamefont
  {J.}~\bibnamefont {Yin}}, \bibinfo {author} {\bibfnamefont {F.}~\bibnamefont
  {Kong}}, \bibinfo {author} {\bibfnamefont {Y.}~\bibnamefont {Lin}}, \bibinfo
  {author} {\bibfnamefont {C.}~\bibnamefont {Yin}}, \bibinfo {author}
  {\bibfnamefont {F.}~\bibnamefont {Shi}}, \bibinfo {author} {\bibfnamefont
  {Y.}~\bibnamefont {Wang}},\ and\ \bibinfo {author} {\bibfnamefont
  {J.}~\bibnamefont {Du}},\ }\bibfield  {title} {\bibinfo {title}
  {{High-fidelity single-shot readout of single electron spin in diamond with
  spin-to-charge conversion}},\ }\href
  {https://doi.org/10.1038/s41467-021-21781-5} {\bibfield  {journal} {\bibinfo
  {journal} {Nature Communications}\ }\textbf {\bibinfo {volume} {12}},\
  \bibinfo {pages} {1} (\bibinfo {year} {2021})}\BibitemShut {NoStop}%
\bibitem [{\citenamefont {Goldman}\ \emph {et~al.}(2015)\citenamefont
  {Goldman}, \citenamefont {Sipahigil}, \citenamefont {Doherty}, \citenamefont
  {Yao}, \citenamefont {Bennett}, \citenamefont {Markham}, \citenamefont
  {Twitchen}, \citenamefont {Manson}, \citenamefont {Kubanek},\ and\
  \citenamefont {Lukin}}]{Goldman2015Phonon-inducedCenters}%
  \BibitemOpen
  \bibfield  {author} {\bibinfo {author} {\bibfnamefont {M.~L.}\ \bibnamefont
  {Goldman}}, \bibinfo {author} {\bibfnamefont {A.}~\bibnamefont {Sipahigil}},
  \bibinfo {author} {\bibfnamefont {M.~W.}\ \bibnamefont {Doherty}}, \bibinfo
  {author} {\bibfnamefont {N.~Y.}\ \bibnamefont {Yao}}, \bibinfo {author}
  {\bibfnamefont {S.~D.}\ \bibnamefont {Bennett}}, \bibinfo {author}
  {\bibfnamefont {M.}~\bibnamefont {Markham}}, \bibinfo {author} {\bibfnamefont
  {D.~J.}\ \bibnamefont {Twitchen}}, \bibinfo {author} {\bibfnamefont {N.~B.}\
  \bibnamefont {Manson}}, \bibinfo {author} {\bibfnamefont {A.}~\bibnamefont
  {Kubanek}},\ and\ \bibinfo {author} {\bibfnamefont {M.~D.}\ \bibnamefont
  {Lukin}},\ }\bibfield  {title} {\bibinfo {title} {{Phonon-induced population
  dynamics and intersystem crossing in nitrogen-vacancy centers}},\ }\href
  {https://doi.org/10.1103/PhysRevLett.114.145502} {\bibfield  {journal}
  {\bibinfo  {journal} {Physical Review Letters}\ }\textbf {\bibinfo {volume}
  {114}},\ \bibinfo {pages} {1} (\bibinfo {year} {2015})}\BibitemShut {NoStop}%
\bibitem [{\citenamefont {Oberg}\ \emph {et~al.}(2019)\citenamefont {Oberg},
  \citenamefont {Huang}, \citenamefont {Reddy}, \citenamefont {Alkauskas},
  \citenamefont {Greentree}, \citenamefont {Cole}, \citenamefont {Manson},
  \citenamefont {Meriles},\ and\ \citenamefont
  {Doherty}}]{Oberg2019SpinDiamond}%
  \BibitemOpen
  \bibfield  {author} {\bibinfo {author} {\bibfnamefont {L.~M.}\ \bibnamefont
  {Oberg}}, \bibinfo {author} {\bibfnamefont {E.}~\bibnamefont {Huang}},
  \bibinfo {author} {\bibfnamefont {P.~M.}\ \bibnamefont {Reddy}}, \bibinfo
  {author} {\bibfnamefont {A.}~\bibnamefont {Alkauskas}}, \bibinfo {author}
  {\bibfnamefont {A.~D.}\ \bibnamefont {Greentree}}, \bibinfo {author}
  {\bibfnamefont {J.~H.}\ \bibnamefont {Cole}}, \bibinfo {author}
  {\bibfnamefont {N.~B.}\ \bibnamefont {Manson}}, \bibinfo {author}
  {\bibfnamefont {C.~A.}\ \bibnamefont {Meriles}},\ and\ \bibinfo {author}
  {\bibfnamefont {M.~W.}\ \bibnamefont {Doherty}},\ }\bibfield  {title}
  {\bibinfo {title} {{Spin coherent quantum transport of electrons between
  defects in diamond}},\ }\href {https://doi.org/10.1515/nanoph-2019-0144}
  {\bibfield  {journal} {\bibinfo  {journal} {Nanophotonics}\ }\textbf
  {\bibinfo {volume} {8}},\ \bibinfo {pages} {1975} (\bibinfo {year}
  {2019})}\BibitemShut {NoStop}%
\bibitem [{\citenamefont {De~Oliveira}\ \emph {et~al.}(2017)\citenamefont
  {De~Oliveira}, \citenamefont {Antonov}, \citenamefont {Wang}, \citenamefont
  {Neumann}, \citenamefont {Momenzadeh}, \citenamefont {H{\"{a}}u{\ss}ermann},
  \citenamefont {Pasquarelli}, \citenamefont {Denisenko},\ and\ \citenamefont
  {Wrachtrup}}]{DeOliveira2017}%
  \BibitemOpen
  \bibfield  {author} {\bibinfo {author} {\bibfnamefont {F.~F.}\ \bibnamefont
  {De~Oliveira}}, \bibinfo {author} {\bibfnamefont {D.}~\bibnamefont
  {Antonov}}, \bibinfo {author} {\bibfnamefont {Y.}~\bibnamefont {Wang}},
  \bibinfo {author} {\bibfnamefont {P.}~\bibnamefont {Neumann}}, \bibinfo
  {author} {\bibfnamefont {S.~A.}\ \bibnamefont {Momenzadeh}}, \bibinfo
  {author} {\bibfnamefont {T.}~\bibnamefont {H{\"{a}}u{\ss}ermann}}, \bibinfo
  {author} {\bibfnamefont {A.}~\bibnamefont {Pasquarelli}}, \bibinfo {author}
  {\bibfnamefont {A.}~\bibnamefont {Denisenko}},\ and\ \bibinfo {author}
  {\bibfnamefont {J.}~\bibnamefont {Wrachtrup}},\ }\bibfield  {title} {\bibinfo
  {title} {{Tailoring spin defects in diamond by lattice charging}},\
  }\bibfield  {journal} {\bibinfo  {journal} {Nature Communications}\ }\textbf
  {\bibinfo {volume} {8}},\ \href {https://doi.org/10.1038/ncomms15409}
  {10.1038/ncomms15409} (\bibinfo {year} {2017})\BibitemShut {NoStop}%
\bibitem [{\citenamefont {Sasama}\ \emph {et~al.}(2020)\citenamefont {Sasama},
  \citenamefont {Kageura}, \citenamefont {Komatsu}, \citenamefont {Moriyama},
  \citenamefont {Inoue}, \citenamefont {Imura}, \citenamefont {Watanabe},
  \citenamefont {Taniguchi}, \citenamefont {Uchihashi},\ and\ \citenamefont
  {Takahide}}]{Sasama2020Charge-carrierTransistors}%
  \BibitemOpen
  \bibfield  {author} {\bibinfo {author} {\bibfnamefont {Y.}~\bibnamefont
  {Sasama}}, \bibinfo {author} {\bibfnamefont {T.}~\bibnamefont {Kageura}},
  \bibinfo {author} {\bibfnamefont {K.}~\bibnamefont {Komatsu}}, \bibinfo
  {author} {\bibfnamefont {S.}~\bibnamefont {Moriyama}}, \bibinfo {author}
  {\bibfnamefont {J.~I.}\ \bibnamefont {Inoue}}, \bibinfo {author}
  {\bibfnamefont {M.}~\bibnamefont {Imura}}, \bibinfo {author} {\bibfnamefont
  {K.}~\bibnamefont {Watanabe}}, \bibinfo {author} {\bibfnamefont
  {T.}~\bibnamefont {Taniguchi}}, \bibinfo {author} {\bibfnamefont
  {T.}~\bibnamefont {Uchihashi}},\ and\ \bibinfo {author} {\bibfnamefont
  {Y.}~\bibnamefont {Takahide}},\ }\bibfield  {title} {\bibinfo {title}
  {{Charge-carrier mobility in hydrogen-terminated diamond field-effect
  transistors}},\ }\bibfield  {journal} {\bibinfo  {journal} {Journal of
  Applied Physics}\ }\textbf {\bibinfo {volume} {127}},\ \href
  {https://doi.org/10.1063/5.0001868} {10.1063/5.0001868} (\bibinfo {year}
  {2020})\BibitemShut {NoStop}%
\bibitem [{\citenamefont {Silchter}(1996)}]{Silchter1996PrinciplesResonance}%
  \BibitemOpen
  \bibfield  {author} {\bibinfo {author} {\bibfnamefont {C.~P.}\ \bibnamefont
  {Silchter}},\ }\href@noop {} {\emph {\bibinfo {title} {Principles of Magnetic
  Resonance}}},\ \bibinfo {edition} {3rd}\ ed.\ (\bibinfo  {publisher}
  {Springer},\ \bibinfo {address} {Heidelberg},\ \bibinfo {year}
  {1996})\BibitemShut {NoStop}%
\bibitem [{\citenamefont {Oberg}\ \emph {et~al.}(2020)\citenamefont {Oberg},
  \citenamefont {De~Vries}, \citenamefont {Hanlon}, \citenamefont {Strazdins},
  \citenamefont {Barson}, \citenamefont {Doherty},\ and\ \citenamefont
  {Wrachtrup}}]{Oberg2020SolutionElectrometers}%
  \BibitemOpen
  \bibfield  {author} {\bibinfo {author} {\bibfnamefont {L.~M.}\ \bibnamefont
  {Oberg}}, \bibinfo {author} {\bibfnamefont {M.~O.}\ \bibnamefont {De~Vries}},
  \bibinfo {author} {\bibfnamefont {L.}~\bibnamefont {Hanlon}}, \bibinfo
  {author} {\bibfnamefont {K.}~\bibnamefont {Strazdins}}, \bibinfo {author}
  {\bibfnamefont {M.~S.}\ \bibnamefont {Barson}}, \bibinfo {author}
  {\bibfnamefont {M.~W.}\ \bibnamefont {Doherty}},\ and\ \bibinfo {author}
  {\bibfnamefont {J.}~\bibnamefont {Wrachtrup}},\ }\bibfield  {title} {\bibinfo
  {title} {{Solution to Electric Field Screening in Diamond Quantum
  Electrometers}},\ }\href {https://doi.org/10.1103/PhysRevApplied.14.014085}
  {\bibfield  {journal} {\bibinfo  {journal} {Physical Review Applied}\
  }\textbf {\bibinfo {volume} {14}},\ \bibinfo {pages} {1} (\bibinfo {year}
  {2020})}\BibitemShut {NoStop}%
\bibitem [{\citenamefont {Stacey}\ \emph {et~al.}(2019)\citenamefont {Stacey},
  \citenamefont {Dontschuk}, \citenamefont {Chou}, \citenamefont {Broadway},
  \citenamefont {Schenk}, \citenamefont {Sear}, \citenamefont {Tetienne},
  \citenamefont {Hoffman}, \citenamefont {Prawer}, \citenamefont {Pakes},
  \citenamefont {Tadich}, \citenamefont {de~Leon}, \citenamefont {Gali},\ and\
  \citenamefont {Hollenberg}}]{Stacey2019EvidenceSources}%
  \BibitemOpen
  \bibfield  {author} {\bibinfo {author} {\bibfnamefont {A.}~\bibnamefont
  {Stacey}}, \bibinfo {author} {\bibfnamefont {N.}~\bibnamefont {Dontschuk}},
  \bibinfo {author} {\bibfnamefont {J.~P.}\ \bibnamefont {Chou}}, \bibinfo
  {author} {\bibfnamefont {D.~A.}\ \bibnamefont {Broadway}}, \bibinfo {author}
  {\bibfnamefont {A.~K.}\ \bibnamefont {Schenk}}, \bibinfo {author}
  {\bibfnamefont {M.~J.}\ \bibnamefont {Sear}}, \bibinfo {author}
  {\bibfnamefont {J.~P.}\ \bibnamefont {Tetienne}}, \bibinfo {author}
  {\bibfnamefont {A.}~\bibnamefont {Hoffman}}, \bibinfo {author} {\bibfnamefont
  {S.}~\bibnamefont {Prawer}}, \bibinfo {author} {\bibfnamefont {C.~I.}\
  \bibnamefont {Pakes}}, \bibinfo {author} {\bibfnamefont {A.}~\bibnamefont
  {Tadich}}, \bibinfo {author} {\bibfnamefont {N.~P.}\ \bibnamefont {de~Leon}},
  \bibinfo {author} {\bibfnamefont {A.}~\bibnamefont {Gali}},\ and\ \bibinfo
  {author} {\bibfnamefont {L.~C.}\ \bibnamefont {Hollenberg}},\ }\bibfield
  {title} {\bibinfo {title} {{Evidence for Primal sp 2 Defects at the Diamond
  Surface: Candidates for Electron Trapping and Noise Sources}},\ }\href
  {https://doi.org/10.1002/admi.201801449} {\bibfield  {journal} {\bibinfo
  {journal} {Advanced Materials Interfaces}\ }\textbf {\bibinfo {volume} {6}},\
  \bibinfo {pages} {1} (\bibinfo {year} {2019})}\BibitemShut {NoStop}%
\bibitem [{\citenamefont {Razinkovas}\ \emph {et~al.}(2021)\citenamefont
  {Razinkovas}, \citenamefont {Doherty}, \citenamefont {Manson}, \citenamefont
  {Van De~Walle},\ and\ \citenamefont
  {Alkauskas}}]{Razinkovas2021VibrationalDiamond}%
  \BibitemOpen
  \bibfield  {author} {\bibinfo {author} {\bibfnamefont {L.}~\bibnamefont
  {Razinkovas}}, \bibinfo {author} {\bibfnamefont {M.~W.}\ \bibnamefont
  {Doherty}}, \bibinfo {author} {\bibfnamefont {N.~B.}\ \bibnamefont {Manson}},
  \bibinfo {author} {\bibfnamefont {C.~G.}\ \bibnamefont {Van De~Walle}},\ and\
  \bibinfo {author} {\bibfnamefont {A.}~\bibnamefont {Alkauskas}},\ }\bibfield
  {title} {\bibinfo {title} {{Vibrational and vibronic structure of isolated
  point defects: The nitrogen-vacancy center in diamond}},\ }\href
  {https://doi.org/10.1103/PhysRevB.104.045303} {\bibfield  {journal} {\bibinfo
   {journal} {Physical Review B}\ }\textbf {\bibinfo {volume} {104}},\ \bibinfo
  {pages} {1} (\bibinfo {year} {2021})}\BibitemShut {NoStop}%
\bibitem [{\citenamefont {Manson}\ \emph {et~al.}(2018)\citenamefont {Manson},
  \citenamefont {Hedges}, \citenamefont {Barson}, \citenamefont {Ahlefeldt},
  \citenamefont {Doherty}, \citenamefont {Abe}, \citenamefont {Ohshima},\ and\
  \citenamefont {Sellars}}]{Manson2018NV--N+Diamond}%
  \BibitemOpen
  \bibfield  {author} {\bibinfo {author} {\bibfnamefont {N.~B.}\ \bibnamefont
  {Manson}}, \bibinfo {author} {\bibfnamefont {M.}~\bibnamefont {Hedges}},
  \bibinfo {author} {\bibfnamefont {M.~S.}\ \bibnamefont {Barson}}, \bibinfo
  {author} {\bibfnamefont {R.}~\bibnamefont {Ahlefeldt}}, \bibinfo {author}
  {\bibfnamefont {M.~W.}\ \bibnamefont {Doherty}}, \bibinfo {author}
  {\bibfnamefont {H.}~\bibnamefont {Abe}}, \bibinfo {author} {\bibfnamefont
  {T.}~\bibnamefont {Ohshima}},\ and\ \bibinfo {author} {\bibfnamefont {M.~J.}\
  \bibnamefont {Sellars}},\ }\bibfield  {title} {\bibinfo {title} {{NV--N+ pair
  centre in 1b diamond}},\ }\bibfield  {journal} {\bibinfo  {journal} {New
  Journal of Physics}\ }\textbf {\bibinfo {volume} {20}},\ \href
  {https://doi.org/10.1088/1367-2630/aaec58} {10.1088/1367-2630/aaec58}
  (\bibinfo {year} {2018})\BibitemShut {NoStop}%
\end{thebibliography}%
